\documentclass[3p]{elsarticle}
\usepackage{amssymb}

\usepackage{graphics}
\usepackage{color}
\usepackage{units}
\usepackage{subfigure}
\usepackage{epstopdf}
\usepackage{amsmath}

\definecolor{darkred}{rgb}{0.5,0,0}
\definecolor{darkgreen}{rgb}{0,0.5,0}
\definecolor{darkblue}{rgb}{0,0,0.5}
\definecolor{violett}{rgb}{0.5,0,0.5}
\definecolor{pink}{rgb}{1,0,1}

\journal{Astroparticle Physics}

\makeatletter
  \renewcommand{\fps@figure}{htb}
  \renewcommand{\fps@table}{htb}
\makeatother

\begin{document}


\begin{frontmatter}

\title{On the sensitivity of the HAWC observatory to gamma-ray bursts}

\author[label1]{A.~U.~Abeysekara}
\author[label2]{J.~A.~Aguilar}
\author[label3]{S.~Aguilar}
\author[label3]{R.~Alfaro}
\author[label3]{E.~Almaraz}
\author[label4]{C.~\'Alvarez}
\author[label5]{J.~de~D.~\'Alvarez-Romero}
\author[label3]{M.~\'Alvarez}
\author[label4]{R.~Arceo}
\author[label5]{J.~C.~Arteaga-Vel\'azquez}
% ...
\cortext[corresponding]{Corresponding author: ignacio.taboada@physics.gatech.edu}
\address[label1]{Department of Physics \& Astronomy, Michigan State
  University, 3245 BPS Building, East Lansing, MI 48824, USA}
\address[label2]{Dept. of Physics, University of Wisconsin - Madison, 1150 University Ave, WI 53706, USA}
\address[label3]{Instituto de F\'isica, Universidad Nacional Aut\'onoma de M\'exico, Apartado Postal 20-364, 01000 M\'exico D.F., M\'exico   }
\address[label4]{CEFYMAP, Universidad Aut\'onoma de Chiapas, 4a. Oriente Norte No. 1428, Col. La pimienta, Tuxtla Guti\'errez, Chiapas, 29040 M\'exico   }
\address[label5]{Universidad Michocana de San Nicol\'as de Hidalgo, Morelia, Mich. 58040, M\'exico   }
% ...

\begin{abstract}
We present the sensitivity of HAWC to Gamma Ray Bursts (GRBs). HAWC is
a very high-energy gamma-ray observatory currently under construction
in Mexico at an altitude of \unit[4100]{m}. It will observe
atmospheric air showers via the water Cherenkov method. HAWC will
consist of 300 large water tanks instrumented with 4 photomultipliers each.  HAWC
has two data acquisition (DAQ) systems. The main DAQ system reads out
coincident signals in the tanks and reconstructs the direction and
energy of individual atmospheric showers. The scaler DAQ counts the
hits in each photomultiplier tube (PMT) in the detector and searches
for a statistical excess  over the noise of all PMTs.  We show that
HAWC has a realistic opportunity to observe the high-energy power law
components of GRBs that extend at least up to 30~GeV, as it has been
observed by Fermi LAT. The two DAQ systems have an energy threshold
that is low enough to observe events similar to GRB~090510 and
GRB~090902b with the characteristics observed by Fermi LAT. HAWC will
provide information about the high-energy spectra of GRBs which in
turn could help to understanding about e-pair attenuation in GRB jets,
extragalactic background light absorption, as well as establishing the
highest energy to which GRBs accelerate particles. 
\end{abstract}


\end{frontmatter}


\section{Introduction}
\label{sec:intro}

Gamma-ray bursts (GRBs) are among the most powerful events in the
universe \cite{gehrels2009,racusin}.
Suggested progenitors for GRBs include
neutron star-neutron star or neutron star-black hole mergers
\cite{janka,rosswog}, millisecond proto-magnetars \cite{thompson} and the core
collapse of massive stars \cite{woosley, macfadyen}. Most of these theories 
have in common that the initial source for the
energy output of the GRB is a central black hole surrounded by the
remnant matter of the progenitor.
A jetted, highly relativistic fireball interacting with itself or the 
surrounding interstellar matter, forming internal and external shocks in 
which Fermi-acceleration takes place, delivers a plausible explanation 
of the non-thermal spectrum of GRBs \cite{meszaros01,meszaros06,piran99}. 
The boosted emission from GRBs explains why high-energy gamma rays are 
not attenuated via e-pair production \cite{lazatti07}. However, the connection 
between accretion and jet production is poorly understood. The high luminosity 
of these events allows for their detection at very high redshifts \cite{tanvir09}, 
making them valuable for answering many astrophysical questions, 
even if their origin remains unclear. In particular, GRBs probe the content of the
intervening space between their origin and Earth. 

Measurements of gamma-ray light curves reveal two classes of
bursts \cite{kouveliotou93}: long and short GRBs, 
if they are longer or shorter than \unit[2]{s} respectively. The 
prompt emission of a gamma-ray burst is typically
described by the \textit{Band function} \cite{band}. The Band function
is a good fit to the majority of GRB prompt
spectra. Compton Gamma Ray Observatory (CGRO) data of GRB~941017 \cite{grb941017} and RHESSI data
of GRB~021206 \cite{grb021206} showed that an additional hard power
law component is sometimes present. 
As of mid-2011, the Fermi Large Area Telescope (Fermi LAT, which
nominally operates between \unit[30]{MeV} and \unit[300]{GeV}) had
detected 26 GRBs. Among these, the long GRB~090902b \cite{grb090902b}
and the short GRB~090510 
\cite{grb090510} are notable for being very bright and having non-Band
hard power-law components. To date, the highest energy photon recorded
from a GRB is \unit[33]{GeV} from GRB~090902b 
(or \unit[94]{GeV} corrected for redshift) \cite{grb090902b}. Fermi LAT probably
did not detect higher energy gamma rays from GRB~090902b because of its limited size. 
Fermi Gamma-ray Burst Monitor (Fermi-GBM, which operates between \unit[8]{keV}
and \unit[40]{MeV}), provided observations of the short bursts GRB~090227B and 
GRB~090228 \cite{FermiSymp} that are also best explained by including an additional hard
non-Band component. Milagrito, predecessor of the Milagro detector,
reported a possible detection of gamma rays in the TeV energy range
from GRB~970417A at $3\, \sigma$ level \cite{atkins03}. Additional
searches by Milagro over 7 years, did not result in significant
observations \cite{milagroGRB1,milagroGRB2,milagroGRB3}. The non-Band
component is currently a challenge to GRB models. Among others, it has
been interpreted as proton synchrotron radiation in the prompt phase
\cite{razzaque} or electron synchrotron radiation in the early
afterglow phase \cite{ghirlanda}.

An observed GRB high-energy spectral cutoff can provide information about the
source itself and the propagation of the gamma rays through the interstellar
media (ISM), as well as insight into aspects of fundamental physics. Source specific
information is, for example, the bulk Lorentz boost factor in a jet model
\cite{rhoads97} or the optical properties in the source volume
\cite{gilmore10}. Direct measurement of the bulk Lorentz boost factor ($\Gamma$) in 
a GRB jet remains elusive. Recently reported observations of a 
spectral cutoff in GRB~090926, however, can be interpreted as a measurement of
$\Gamma$ \cite{grb090926}. For other GRBs only a lower limit is
available, e.g. for GRB~090510 $\Gamma \gtrsim 1000$
\cite{grb090510}. During the
propagation of gamma rays through the ISM, an attenuation due to interactions with the
extra-galactic background light (EBL) is expected \cite{gilmore09}.
Consequently, the spectral energy cutoff can be a probe for the EBL density.
Measurements deliver a relatively high cutoff energy compared to current EBL
model predictions \cite{magic08}. This can be an indicator for physics beyond
the standard model \cite{sanchez09}. The broad energy range in GRB spectra and
the prompt emission, especially in short GRBs, allow for the measurement of
bulk Lorentz invariance violation \cite{biesiada09,abdo09a}. 

Because of the important information high-energy gamma rays are able to provide,
many instruments for their detection have been installed. Currently three major
classes of high-energy detectors exist: Satellite detectors (e.g.
\cite{gehrels94,thompson05,atwood09,meegan09,gehrels04}), Imaging Atmospheric
Cherenkov Telescopes (IACTs) \cite{hinton09} and Extensive Air Shower (EAS)
particle detector arrays \cite{sinnis}. Satellites can observe very wide fields
of view (e.g. \unit[2.4]{sr} or 19\% of \unit[$4\pi$]{sr} for Fermi LAT) and have close to a 100\%
operational duty cycle. On the other hand, the limited physical size of
satellites prevents them from obtaining enough statistics to reach energies
greater than tens of GeV. Operating above $\approx \unit[50]{GeV}$ IACTs have
superb sensitivity and angular and energy resolution. However, IACTs can observe
GRBs only in good weather and on moonless nights ($\approx$ 10\% duty cycle for
GRBs), and their field of view is restricted to 5 degrees in diameter
or less. Therefore the crucial prompt observations may not
be possible even with IACTs that have 
been designed for fast slewing (\unit[$\sim$1]{min}).
EAS detector
arrays, such as HAWC, benefit from a very large field of view
($\approx \unit[2]{sr}$ or 16\% of \unit[$4\pi$]{sr}) and near 100\% duty cycle that will allow
for observations in the prompt phase. They are also sensitive to
energies beyond those covered by satellites. EAS observatories, in
particular HAWC, are 
thus useful high-energy GRB detectors that complement the observations by
satellites such as Fermi. In this paper we will present the sensitivity and
capabilities of two methods of detection of GRBs by HAWC and show the
observatory's ability to measure possible high-energy emission from GRBs.

\section{HAWC}
\label{sec:hawc}

\begin{figure*}
  \centering
  \subfigure[HAWC tank layout.]
  {
    \label{fig:layout_detail}
    \includegraphics[width=0.45\linewidth]{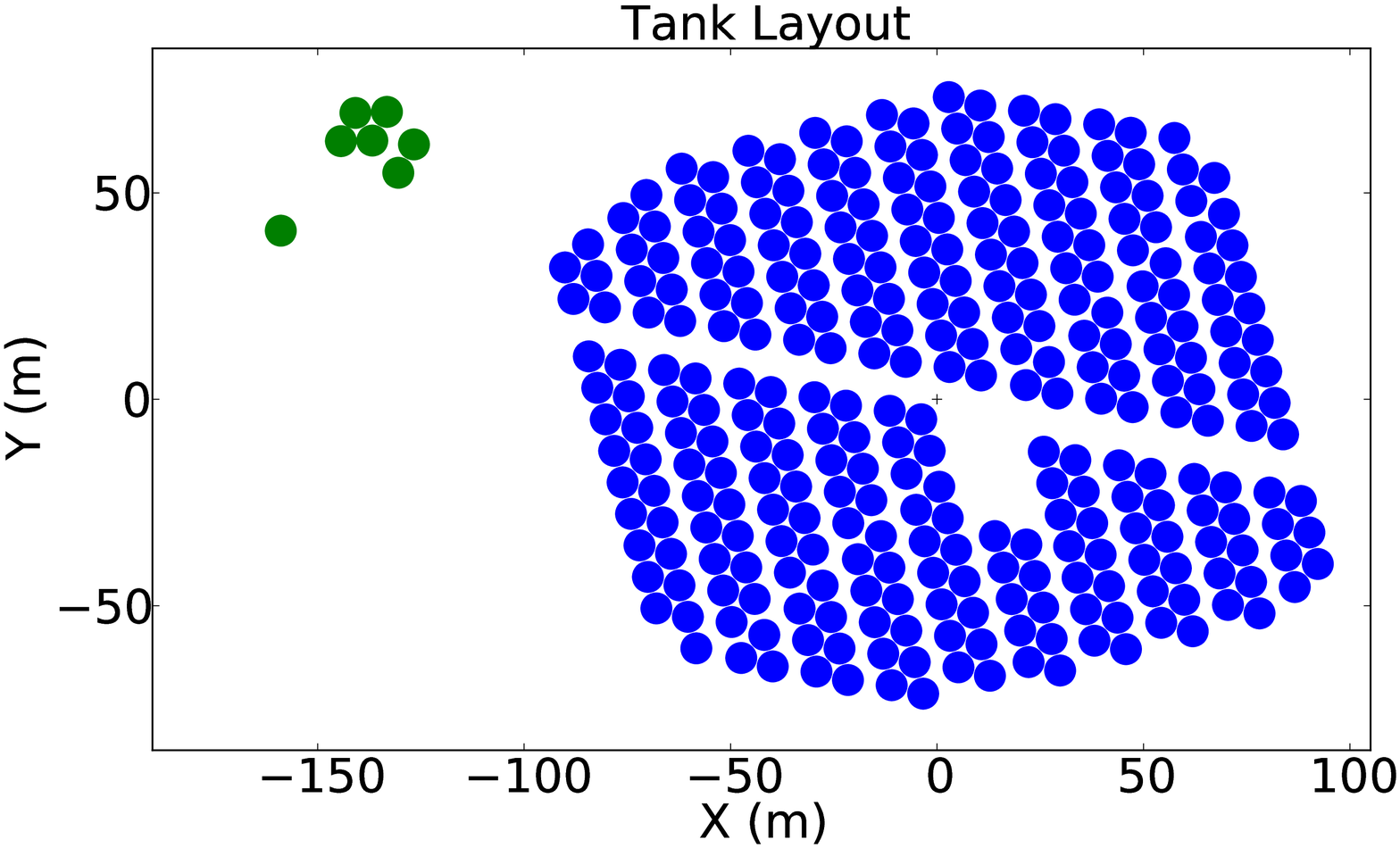}
  }
  \subfigure[Water Cherenkov Detection Principle.]
  {
    \label{fig:tank}
    \includegraphics[width=0.35\linewidth]{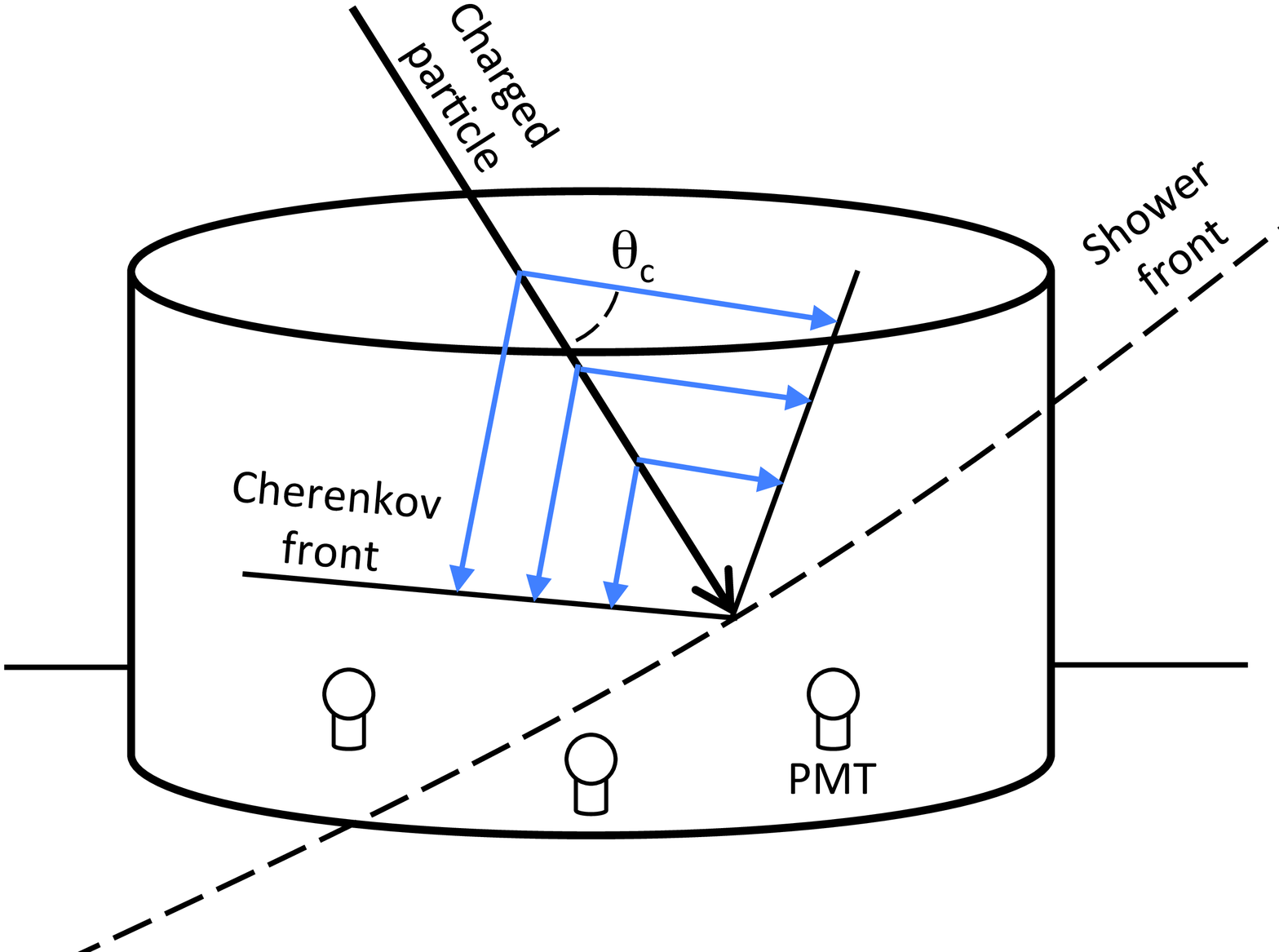}
  }
  \caption[HAWC layout and operation principle.]
          {{\bf HAWC layout and operation principle.} The left panel
            shows the relative position of HAWC tanks. The seven tanks
            at the top left correspond to VAMOS. The electronics
            counting house will be at the empty region in the center
            of the array. The right panel shows the principle of
            Water Cherenkov Detection. Particles, part of an air
            shower, arrive at the ground in a shower
            front. Relativistic charged particles produce Cherenkov
            radiation as they travel in the water tanks. Cherenkov
            radiation is emitted at a precise angle $\theta_c$ with
            respect to the particle trajectory. Cherenkov radiation 
            is detected by photomultiplier tubes at the bottom of the
            tank.} 
  \label{fig:layout}
\end{figure*}

The High Altitude Water Cherenkov (HAWC) observatory is a very
high-energy (VHE) gamma-ray detector currently under construction near
the peak of Volc\'an Sierra Negra, Mexico. HAWC is located at \unit[4100]{m} of altitude, N
$18^\circ 59^\prime 48^{\prime \prime}$, W $97^\circ 18^\prime
34^{\prime \prime}$. When completed in 2014, HAWC will consist of 300
steel tanks of $\unit[7.3]{m}$ diameter and $\unit[4.5]{m}$ deep,
covering an instrumented area of about $\unit[22,000]{m^2}$ (the actual tank
coverage is $\unit[12,550]{m^2}$). Each tank will hold a bladder filled with purified 
water and will contain three $\unit[20]{cm}$ photomultiplier tubes
(PMTs). The PMTs are placed near the bottom of the tank looking up in order
to measure prompt Cherenkov light. The inner walls of the bladders are
dark to reduce reflections of light. 
An additional \unit[25]{cm}, high quantum efficiency PMT will be added
to the center of each tank. However, results presented here correspond
to simulations of three \unit[20]{cm} PMTs per tank.
%
%
The additional PMT will extend HAWC's low 
energy threshold, improving upon what is presented here. A test array of
seven tanks, called VAMOS (Verification And Measuring of Observatory
System), has already been built on site. Six of the tanks have been
filled with water and instrumented with 4 to 7 PMTs per
tank. Engineering data has been collected with 6 tanks. Continuous
operation of VAMOS started in Sept 29, 2011. Operation of the first 30
HAWC tanks is expected to start in 2012. A layout of HAWC and VAMOS as
well as a description of the water Cherenkov detection method can be
seen in figure \ref{fig:layout}.

HAWC observes gamma rays by detecting, at ground level, the particles
that compose an extensive air shower. Charged particles moving through  
water in the tanks generate Cherenkov light that is captured by the
PMTs. Energetic photons traveling through the water in the tanks will
typically Compton scatter or produce an electron-positron pair,
resulting in Cherenkov light. This latter fact is an advantage of the
water Cherenkov method because a large fraction of the electromagnetic
component of an air shower at ground level are photons \cite{sinnis}.

HAWC improves the sensitivity for a Crab-like point spectrum by a factor of
15 in comparison to its predecesor, Milagro \cite{milagro} while also
extending the reach in the low 
 energy region. The trigger in
Milagro used the upper pond layer of $\unit[4,000]{m^2}$, while HAWC
uses its entire instrumented 
area of $\unit[22,000]{m^2}$. For the
purposes of discriminating gamma rays from hadrons, Milagro used its
deep pond layer of $\unit[2,000]{m^2}$, while HAWC can use its entire 
instrumented area of $\unit[22,000]{m^2}$. 
Discrimination of gamma rays
and hadrons is also better in HAWC with respect to Milagro because
detection elements are optically isolated (tanks vs. single
pond). Milagro was complemented by a sparse outrigger array that
extended to about $\unit[40,000]{m^2}$ to improve reconstruction
capabilities. This is not as necessary in HAWC, as the array is 
already big enough to provide excellent reconstruction. Finally the
higher altitude of HAWC ($\unit[4100]{m}$ vs $\unit[2630]{m}$) implies
that the detector is closer to the air shower maximum and for a given
species of primary, more particles are available at ground level. This is
particularly important for the low-energy gamma rays relevant for GRB
observations. HAWC will also be 
able to send quasi-real time alerts (e.g. via the GRB Coordinate
Network, or GCN \cite{gcn}) that can trigger multi-wavelength
campaigns. The VERITAS IACT is geographically located close to HAWC,
and alerts issued by HAWC may be followed by VERITAS.

HAWC data will be collected by two data acquisition systems (DAQs). The
main DAQ will measure the arrival time and time over threshold (TOT) of PMT
pulses, hence providing information for the reconstruction of the
shower core, direction and lateral distribution, which in turn helps to determine the
species of primary
particle and its energy. A secondary DAQ, the scaler 
system, operates in a PMT pulse counting mode \cite{Vernetto} and is
sensitive to gamma ray and cosmic ray (i.e. due to Solar activity)
transient events that produce a sudden increase or decrease in the
counting rates with respect to those produced by atmospheric showers and noise. 

\section{The main DAQ}
\label{sec:triggerdaq}

HAWC's primary DAQ system will record individual events caused by air showers large
enough to simultaneously illuminate a significant fraction of the HAWC array. In
the simplest approach, depending on the number of hit PMTs during a given time
window (trigger condition), a trigger will be issued and sent to time to digital 
converters (TDCs). The TDCs will store the measured times of the PMT hits closest to
the trigger time. The data of each issued trigger are called an
event. For the operation of HAWC we plan to use CAEN VX1190 VME
TDCs. The final triggering configuration of HAWC is still not
defined. As will be shown below, small events contribute significantly
to the sensitivity to GRBs.

The event data recorded by the main DAQ system will consist of the leading and
trailing edges of discriminated PMT pulses. The
Milagro PMTs and front-end boards are being reused in HAWC. 
The TDCs will measure a pulse's leading and trailing edge at two
discriminator settings ($\approx$ 1/4 and $\approx$ 5
photoelectrons). These measurements provide accurate shape to the pulse 
widths, or TOTs, which can be used to measure the pulse charge over a large dynamic
range. The leading and trailing edges will be recorded for two different
discriminator thresholds with $\approx$ \unit[0.5]{ns} accuracy.
Simulations show that an accuracy of \unit[1]{ns} or better is needed to
achieve the best possible angular resolution. Individual events will
be time stamped with a GPS clock, with at least \unit[5]{$\mu$s}
accuracy. This time stamping will allow HAWC to produce a  lightcurve
and measure the variability time for GRBs at very high  energies. 
Data collected by the main DAQ will be passed to an online processing computer farm.
The location of the air shower core on the ground can be estimated from the spatial
distribution of PMT charges and the direction calculated based on the times
of the PMT hits produced as the shower front sweeps across the array. The energy
of the shower can be estimated from number of PMTs hit, and
hadron-induced and photon-induced air showers can be distinguished from each
other on a statistical basis by searching for isolated high-amplitude
pulses \unit[40]{m} or more away from the shower core, indicative of muons in hadronic showers. A
system to send quasi real time alerts in response to main DAQ
gamma-ray transients, e.g. GCN notices, is part of the planned
operations of HAWC.

In HAWC, sources of PMT noise are uncorrelated hits from ambient
radioactivity in the water and in materials composing PMTs and tanks as well as
dark noise from the PMTs. Correlated sources of noise in PMTs, causing several PMTs 
to fire simultaneously, are secondary
gamma rays, electrons and muons from low-energy hadronic cosmic ray
showers. First measurements indicate that the total noise hit rate in each PMT
is $\approx \unit[20]{kHz}$.
The predicted trigger rate is around $\unit[5-20]{kHz}$, mainly
limited by the bandwidth of the TDCs. Examples used in this paper are
$\approx$ 5 and $\approx$ \unit[17]{kHz}. We envision a DAQ system with a deadtime of 1\%
or less.

\subsection{The trigger system}

In the current design, the trigger system is a distinct piece of hardware
that builds a trigger condition based on the number of signals arriving from the
front-end boards. This is referred to as a Simple Multiplicity Trigger (SMT).
If a trigger condition is satisfied, the trigger signals
the TDC to store the data of the PMT hits around the trigger time in a
predefined window.
An alternative solution is also being investigated that combines a 
high-throughput TDC readout with a software trigger.

The vast majority of triggers will be produced by hadronic cosmic ray air
showers that dominate photon triggers by several orders of magnitude. When a
trigger is issued, the data from all PMTs in the array over a time span of
\unit[1-2]{$\mu$s} will be recorded.
The data will be processed online by a dedicated computing farm,
which includes shower reconstruction algorithms and discrimination
of gamma rays from hadrons. 

In order to suppress the PMT noise while keeping
a high efficiency for shower detection, the DAQ can operate under a set of
simple multiplicity triggers with different thresholds. 
The SMT has two parameters: a minimum
number of PMTs above threshold ($nHit$)
and a coincidence window ($\Delta t_{trig}$). 
For vertical showers, the particles arrive at HAWC almost simultaneously,
while for inclined events they are spread out in time.
Hence small values of $\Delta t_{trig}$ guarantee a high efficiency for
vertical showers, while longer 
windows are required for high efficiency at large angles.
Integrating over all angles, air shower particles arrive at HAWC in a
time window of \unit[600]{ns} or less.  In this paper we investigate the
effect of the two triggers mentioned above on the sensitivity to GRBs. A choice of $nHit =
70$ and $\Delta t_{trig} = \unit[190]{ns}$ leads to $\approx$ \unit[5]{kHz}
trigger rate. For $nHit = 30$, $\Delta t_{trig} = \unit[190]{ns}$,  the
rate will be $\approx$ \unit[17]{kHz}.

\subsection{The main DAQ simulation and expected performance of the detector}
\label{sec:simu}

The software employed by Milagro has been modified to simulate
HAWC. 
Galactic cosmic rays are simulated with CORSIKA 6.9 \cite{corsika} for
multiple species with an 
$E^{-2}$ spectrum: protons, He, C, O, Ne, Mg, Si and Fe. The galactic
cosmic ray spectrum is reweighted to match measurements by ATIC
\cite{ATIC}. Gamma-ray showers are also simulated using this machinery
and reweighted for various source spectra, including
power-law spectra as appropriate
to describe the high-energy emission of GRBs. The detector response
model developed for Milagro and modified for HAWC is used at an
altitude of \unit[4100]{m} using a GEANT-4 based code \cite{geant4}. The simulation used for the main DAQ
injects uncorrelated random noise (Poissonian noise) at a rate of
\unit[20]{kHz} per PMT.  

The signal rate $S$ is given by
\begin{equation}
  S(\theta) = \int \textup{d}E\;
              \frac{\textup{d}N}{\textup{d}E}\;
              A_{eff}^{trig}(E,\theta),
\end{equation} where $\textup{d}N/\textup{d}E$
is the photon spectrum and $A_{eff}^{trig}$ is the detector effective area.
$A_{eff}^{trig}$ depends on several variables; here only energy $E$ and zenith
angle $\theta$ are treated.

\begin{figure*}
  \centering
  \subfigure[Effective area of the HAWC for $nHit > 30$.]
 {
    \label{fig:gammaAEff30}
 \includegraphics[width=0.45\linewidth]{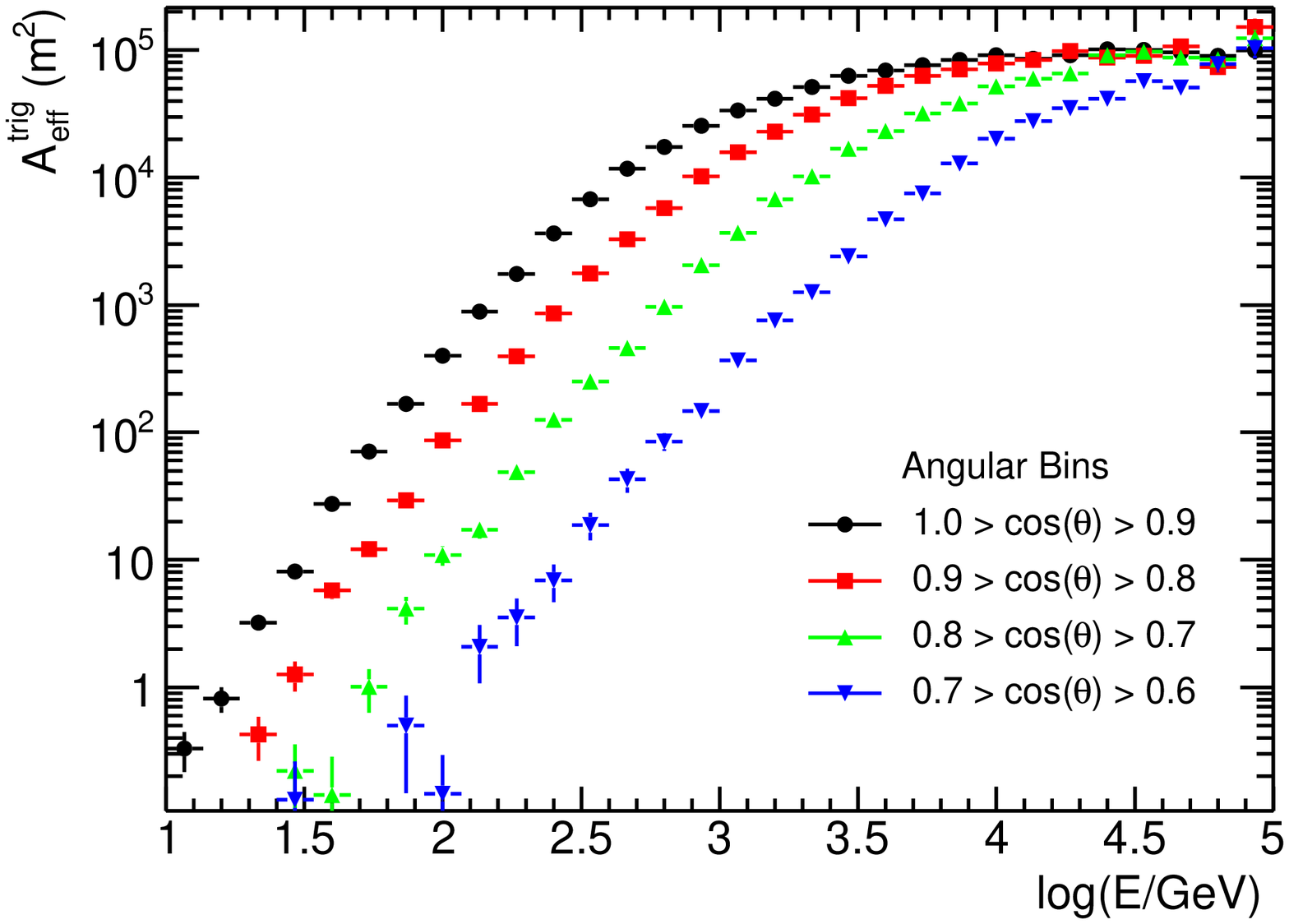}
  }
  \subfigure[Effective area of the HAWC for $nHit > 70$.]
  {
    \label{fig:gammaAEff70}
    \includegraphics[width=0.45\linewidth]{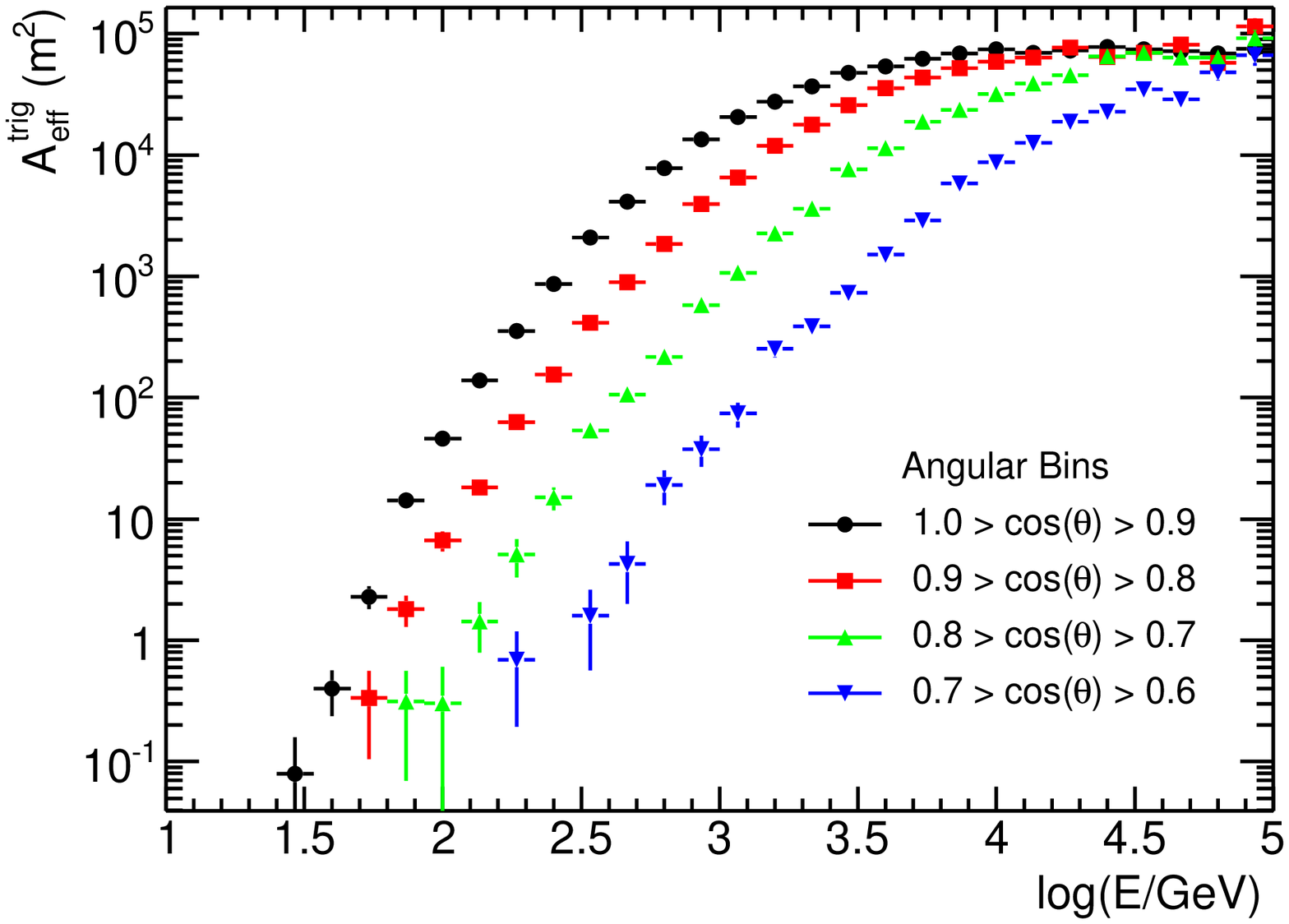}
  }
  \caption[Sensitivity of HAWC using the main DAQ as a function of
  zenith angle.]
          {{\bf Effective area of HAWC using the main DAQ system.}
            Both panels show the effective area $A_{eff}^{trig}$ of
            HAWC in the triggered mode as a function of
            $\gamma$-ray energy for 4 ranges of zenith
            angle. A trigger rate of $\approx \unit[17]{kHz}$ ($nHit >
            $30) is assumed in the left panel. A trigger rate of
            $\approx \unit[5]{kHz}$ ($nHit>$70) is assumed in the
            right panel. Showers reconstructed with $>1.1^\circ$ error
            are excluded for the left panel and $>0.8^\circ$ for the
            right panel. No gamma-hadron separation cut is
            applied. For the energies relevant to GRB searches,
            i.e. below $\approx$300~GeV, applying a gamma-hadron
            separation results in a global reduction of the effective
            area by a factor of 0.85 (left) and 0.75 (right).}  
 \label{fig:gammaAEff}
\end{figure*}

The effective area of HAWC for gamma rays as a function of trigger 
level is shown in Fig.~\ref{fig:gammaAEff}. Although attenuation of
VHE gamma rays via pair production on extragalactic background light
(EBL) will limit the energies of gamma rays from GRBs to be below
\unit[50-300]{GeV} (depending on redshift) \cite{gilmore09}, HAWC retains square meters
to tens of square meters of effective area at these energies, values higher
than Fermi LAT's
$\unit[0.8]{m^2}$. 

\subsection{Sensitivity of the triggered system to GRBs}
\label{triggered}

The sensitivity of HAWC to GRBs depends on a number of factors,
including the GRB emission time scale, emission spectrum, elevation
and redshift, as well as on the trigger, reconstruction and background rejection
capabilities of the experiment.
To calculate HAWC's sensitivity,
we simulate a gamma ray spectrum according to the power-law
$\textup{d}N/\textup{d}E \propto E^{-2}$ with an arbitrary reference flux
normalization. This injection spectrum can be weighted for any other
spectral shapes. In those instances in which we take into account
attenuation of VHE gamma rays due to interaction with extragalactic
background light, the Gilmore et al. model \cite{gilmore09} is used.

After reconstructing the shower core and direction, a hadronic background
rejection cut is applied. The background rejection method is
based on a quantity called `compactness', defined here as the ratio of the
number of PMTs hit in the event to the largest charge on a single PMT
hit at a distance of at least \unit[40]{m} from the reconstructed shower 
core \cite{MilagroCrab}. Hadronic air showers are both clumpy
and muon-rich, and thus tend to exhibit large-amplitude hits at
significant distances from the shower axis. On the other hand
gamma-ray air showers are muon-poor and hence exhibit a more uniform
charge distribution that decreases with distance from the shower
core.
After applying the gamma-hadron separation cut  the gamma ray
efficiency is reduced by $\approx$\,25\% for $nHit>70$ and 
$\approx$\,15\% for $nHit>30$. After applying the gamma-hadron
separtion cut, the background is reduced by $\approx$\,90\% for $nHit>70$  and
$\approx$\,70\% for $nHit>30$.
A cut is also applied to the
angular distance between the reconstructed shower direction and
position of the source. This implies that the GRB position is known
from other observations. The time and duration of the burst are also
assumed known, which allows one to efficiently reject the background
by defining a restrictive time window.

After all the cuts are applied, the rate of background events
remaining is used to determine the minimum number of gamma rays
detectable at $5\, \sigma$ significance. Due to the paucity of the
number of events involved, the $5\, \sigma$ level is defined using
the cumulative Poisson distribution function. The $5\, \sigma$
discovery potential is defined as the flux level which leads to a
50\% probability of detecting a $5\, \sigma$ excess. Using this
definition implies that for very short bursts (\unit[$T\ll1$]{s}) the
sensitivity scales as $1/T$, while for very long bursts (\unit[$T >
3\times10^2$]{s}) the sensitivity scales as $1/\sqrt{T}$. Both the
angular distance cut and the hadron rejection cut are optimized to
maximize the discovery potential. 

\begin{figure}
  \centering
  \includegraphics[width=0.5\linewidth]{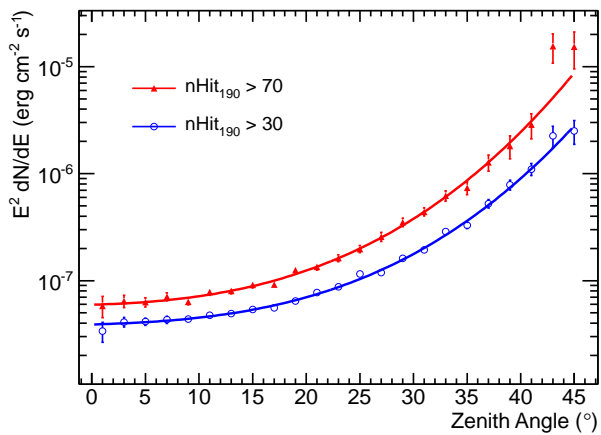}
  \caption[Sensitivity of HAWC using the main DAQ as a function of zenith angle.]
    {{\bf Sensitivity of HAWC using the main DAQ as a function of zenith angle.}
The sensitivity is defined as the flux detectable at $5\, \sigma$ significance with 50\% probability.
Results are given for the baseline trigger ($nHit >$ 70) and a reduced
threshold trigger ($nHit > 30$) for a range of zenith angles of an astrophysical
source. The simulated burst has a duration of $\unit[20]{s}$,  a
spectral index of $-2$ and a redshift of 0.5. EBL attenuation is modeled
following Gilmore et al. \cite{gilmore09}.}
 \label{fig:zenithAng}
\end{figure}

Figure \ref{fig:zenithAng} displays HAWC's sensitivity
as a function of zenith angle for a $\unit[20]{s}$ GRB with an
$E^{-2}$ spectrum at a redshift of $z=0.5$ for two different trigger
thresholds. As can be seen, the low trigger threshold greatly aids in
the detection of GRBs.  The background rate within the angle distance
cut circle  can be parametrized as $R = a \, \exp(b x +  c x^2)$,
where $x = \cos\theta$, and $\theta$ is the zenith angle. After
applying the gamma-hadron separation optimized for GRB search, we find  
that for $nHit>70$ ($nHit>30$), the values of the constants are
 $a=0.42$~Hz ($a=8$~Hz), $b=7.0$ ($b=6.6$) and $c=-12.9$ ($c=-8.6$). 

\begin{figure*}
  \centering
  \subfigure[$\approx$ 5~\unit{kHz} trigger rate ($nHit > 70$)]{
    \includegraphics[width=0.45\linewidth]{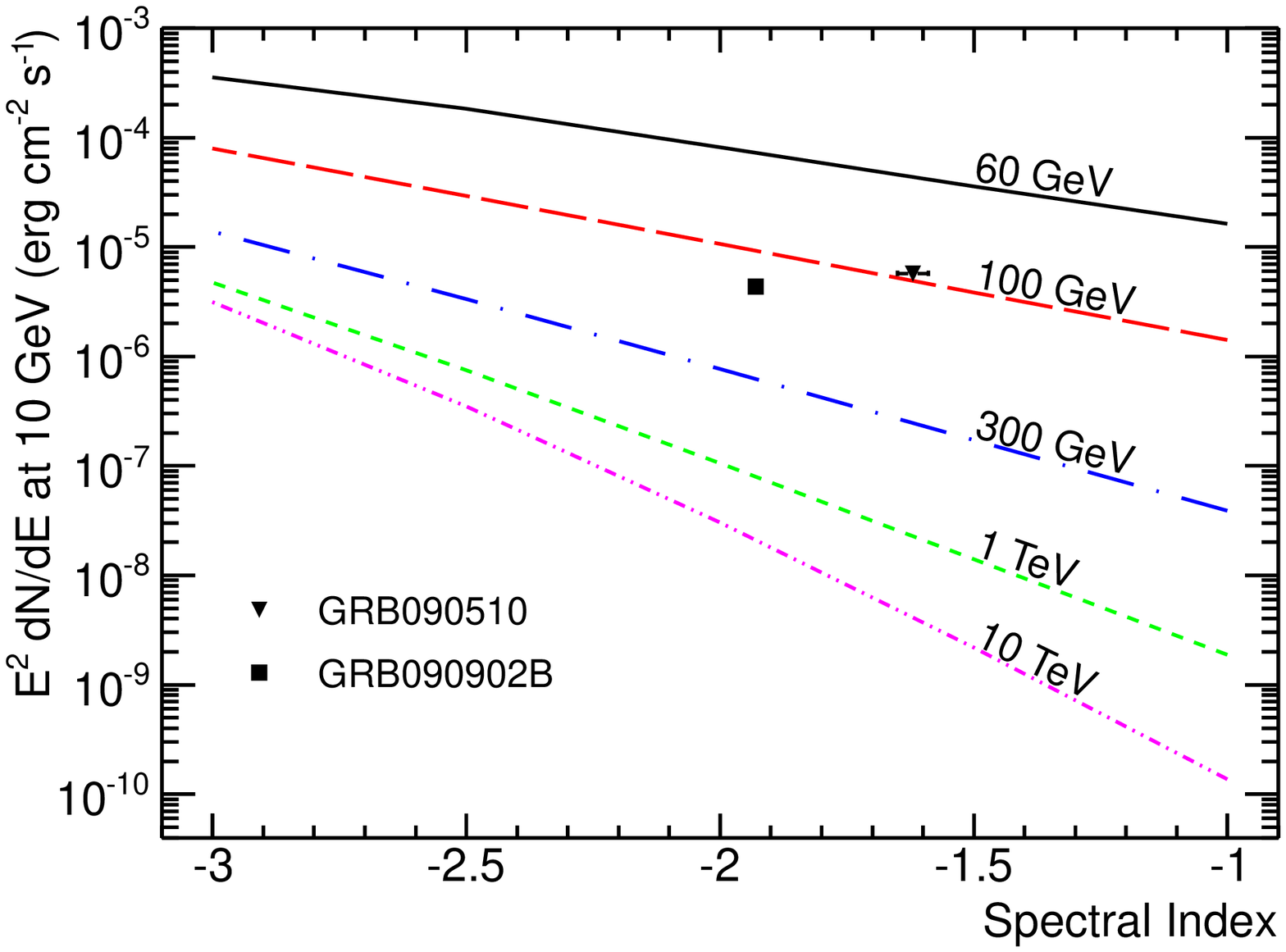}
    \label{P:TrigFermiFluence0}
  }
  \subfigure[$\approx$ 17~\unit{kHz} trigger rate ($nHi t> 30$)]{
   \includegraphics[width=0.45\linewidth]{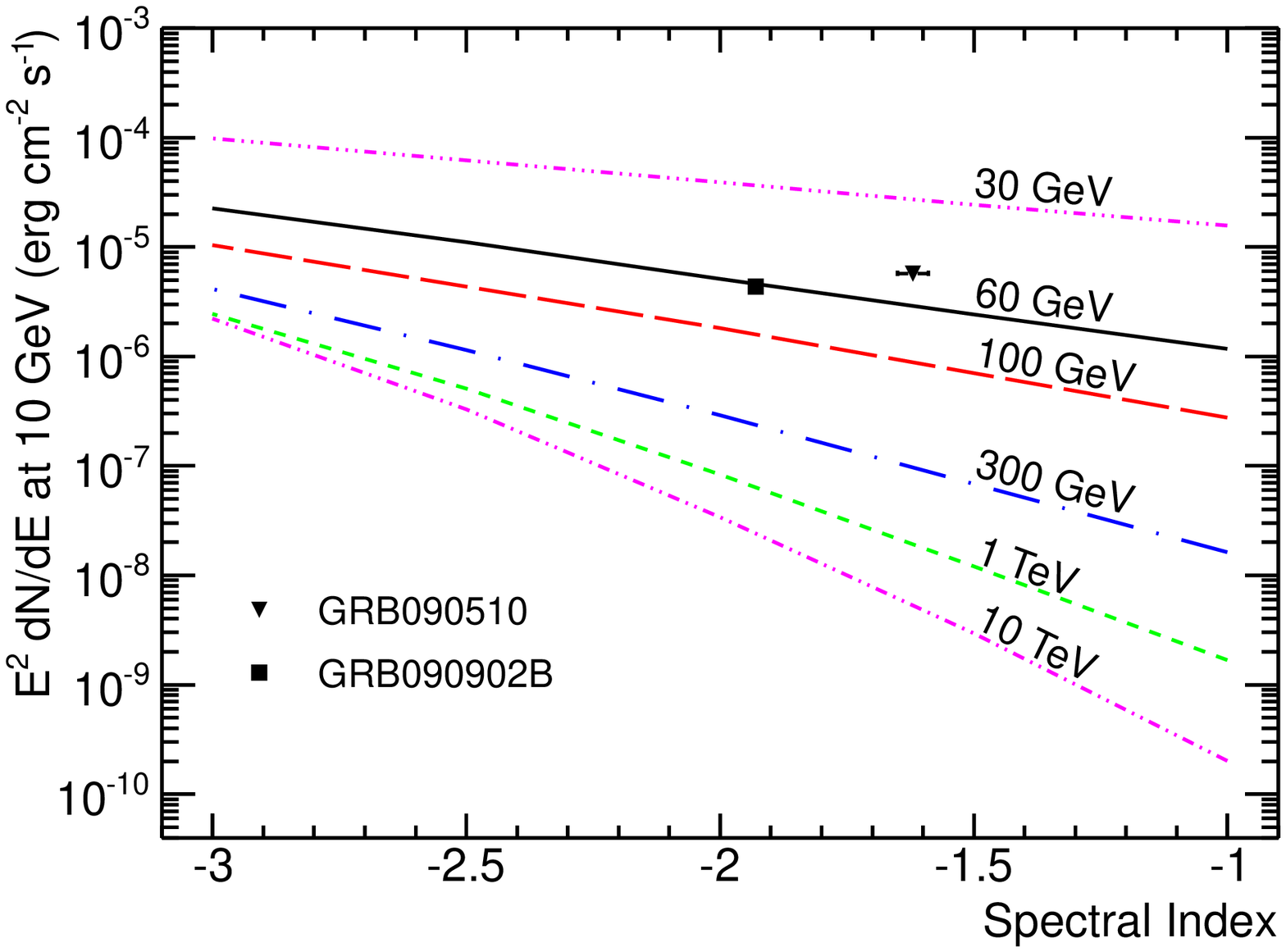}
    \label{P:TrigFermiFluence2}
  }
  \caption[Sensitivity using the main DAQ to Fermi events]
    {{\bf Sensitivity using the main DAQ as a function of spectral index for two different trigger thresholds.}
 The $5\, \sigma$ discovery potential is shown as a function of spectral index
 for various values of a sharp high-energy spectral cutoff.
 The left plot is for the baseline trigger ($nHit >$ 70).
 The right plot is for an alternative low threshold trigger ($nHit >$ 30).
 The duration of the burst is fixed to $\unit[1]{s}$ and the zenith angle is fixed to $20^{\circ}$.
 Data from 2 different GRBs are corrected for duration and inserted for comparison \cite{grb090510,grb090902b}.
 }
  \label{fig:TrigFermiFluence}
\end{figure*}

Figure~\ref{fig:TrigFermiFluence} illustrates the effects of
different GRB emission spectra on the expected sensitivity of HAWC using
the main DAQ. The two panels of Fig.~\ref{fig:TrigFermiFluence} show the 
sensitivity curves for two different trigger thresholds assuming that the burst 
occurs at a zenith angle of $20^{\circ}$ and lasts 1 second. We consider a range 
of spectral indices for spectra of
the type $\textup{d}N/\textup{d}E \propto E^{\gamma}$ with various high-energy
cutoffs. The effect of EBL is not directly considered because it can be
simplistically simulated as a sharp cutoff. As an example, for a
redshift of $z=1$, Gilmore et al. \cite{gilmore09} predict a cutoff at about
$\unit[125]{GeV}$. This choice of spectra is motivated by Fermi LAT
observation of GRBs at high-energy \cite{grb090902b,grb090510}.
Data for GRBs 090510 and 090902b, extracted from \cite{grb090510} and \cite{grb090902b}, are shown for
comparison. The reported fluxes of those bursts were scaled by
$T^{0.7}$, where $T$ is 0.5 and 30 (seconds) for GRB~090510 and GRB~090902b
respectively, to account for the dependence of HAWC's sensitivity on burst
duration as explained previously.
HAWC will be able to detect bursts such as GRB~090510 or GRB~090902b
with high significance if the high-energy cutoff is above $\approx$ \unit[100]{GeV}.
In the configuration with the low trigger threshold ($nHit >\,$30)
cutoff values down to $\approx$ \unit[60]{GeV} would be reachable.

The present estimates of HAWC sensitivity are based on a simple event counting
approach that does not distinguish between low and high energy events. This is
equivalent to a binned analysis with a single energy bin. This approach is valid
for short transients because of the relatively small number of background events
in the search time window. However, for long transients the background
contamination becomes important. This is especially true at low energy, where
both the gamma-hadron separation and angular reconstruction performance degrade.
The resulting energy dependence of the signal-to-noise ratio can be accounted
for by defining several energy bins or via an unbinned (likelihood-based)
analysis. This should improve the sensitivity upon the baseline values presented
here.

The above studies required prior information on a GRB.  A simple way to
estimate the sensitivity in the case of no prior knowledge of a GRB is to divide the sky
and discretize time into bins and apply the technique described above
to each bin. This leads to a large trial factor which has to
be compensated for by requiring more events.
The division of a \unit[2]{sr} field of view into bins of $0.7^{\circ}$
radius leads to $\approx 10^4$ bins, or $10^4$ trial factor per time
bin. Time discretization depends on assumed GRB duration. For
example, dividing \unit[1]{year} of data into \unit[1]{s} intervals leads to a trial
factor of $3\times 10^7$. So the total trial factor is greater than $3\times
10^{11}$ because of oversampling. To make a 5\,$\sigma$ level observation 
the pre-trial factor observations would need to be $\approx$\,8.5\ $\sigma$ (p-value of
$1.8\times 10^{-18}$). This corresponds to a loss in sensitivity of
less than a factor of 2. For example, for a 1 second burst 13 events
would be needed instead of 7 when an external trigger such as GCN is used.

\section{The scaler DAQ}
\label{sec:scalers}

Ground based air shower arrays are sensitive to gamma-ray transients,
GRBs in particular, in the GeV-TeV range using the single particle or
scaler technique \cite{Vernetto}. During regular operations, the
counts in HAWC's PMTs are due to cosmic ray air showers, naturally
occurring radioactivity near the PMTs and thermal noise in the
PMTs. This system can be used to monitor the health of the detector
and to study particle emission by the Sun \cite{SunMilagro}, such as protons, neutrons and
ions. By monitoring PMTs at low threshold ($>1/4$ photoelectron), the
scaler system is sensitive to gamma-ray transients. While the
scaler system is not able to provide directional information, it is
complementary to the trigger system and to other detectors, such as
Fermi LAT, 
that are sensitive to gamma-ray transients. For the operation of HAWC
we plan to use Struck SiS-3820 VME scalers.

All 900 PMT rates will be monitored in $\unit[10]{ms}$ windows.  This
fine time sampling will allow the scaler system to produce a
lightcurve and measure the variability of GRBs at very high
energies. As described before, each PMT is expected to have a rate of
$\approx \unit[20]{kHz}$. A 
transient flux of gamma rays will result in a detector wide increase
of the PMT rate, thus gamma-ray transients are identified on a
statistical basis. Low energy gamma rays from GRBs that are not
observed by the main DAQ may still be observable by the scaler
system. Issuing quasi-real time alerts is not being considered as part
of the default operation of HAWC scalers, but we will distribute GCN
circulars describing HAWC scaler observations of GRBs as appropriate.

\subsection{Signal Simulation for the scaler DAQ}
\label{sec:signalscaler}

\begin{figure}
  \centering
  \includegraphics[width=0.5\linewidth]{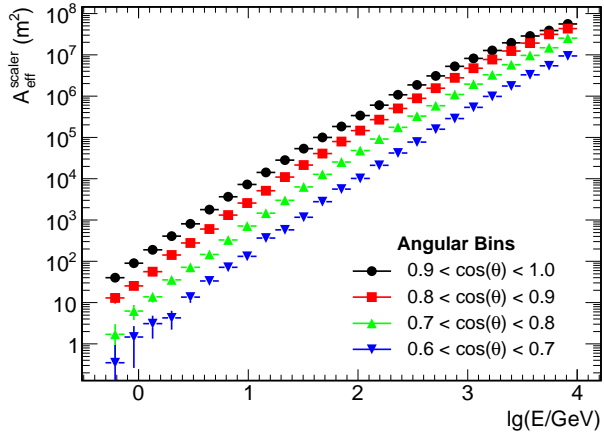}
  \caption[Effective area of the HAWC scaler system.]
    {{\bf Effective area of the HAWC scaler system.}  The scaler
      effective area $A_{eff}^{scaler}$ is shown for 4 zenith angle bins.}
  \label{fig:ScalerEffArea}
\end{figure}

Signal simulation has been performed with the same software as the
main DAQ. The energy range for primary showers was set to $\unit[0.5]{GeV}$ -
$\unit[10]{TeV}$, because we expect the scalers to have sensitivity to
lower energy gamma rays than the main DAQ.  Photons of energy as low
as a few GeV can have a significant contribution to the scaler
signal. The signal rate $S$ in the scaler
system is the number of PMT counts expected on excess of the nominal
PMT rate. This signal rate $S$ is given by
\begin{equation}
S(\theta) = \int \textup{d}E \frac{\textup{d}N}{\textup{d}E}\;
            A_{eff}^{scaler}(E,\theta), 
\end{equation}
where $\textup{d}N/\textup{d}E$ is the photon spectrum and $A_{eff}^{scaler}$ is
the effective area for the scaler system. Because each atmospheric
shower can cause hits in multiple PMTs, the scaler effective area is:
\begin{equation}
A_{eff}^{scaler}(E,\theta) = A_{thrown} N_{PMT}\frac{N_{obs}(E,\theta)}{N_{thrown}(E,\theta)},
\end{equation}
where $A_{thrown}$ is the area over which simulated events are 
thrown, $N_{obs}$ is the number of showers that result in at least one PMT being
hit, $N_{thrown}$ is the number of simulated showers and $N_{PMT}$ is the
average number of PMTs hit for energy $E$ and zenith
$\theta$. Because the scaler DAQ signal is the number of PMT 
hits and not individual air showers, the value for the effective area
is not restricted to the physical 
detector size. We have used the same simulation as the main DAQ system to
calculate the scaler effective area $A_{eff}^{scaler}$ of HAWC. The scaler
effective area as a function of energy for various zenith bands is shown in
Fig.~\ref{fig:ScalerEffArea}. Besides simulating HAWC with 300 tanks and 3 PMTs
per tank as described in Sec. \ref{sec:hawc}, we have also simulated
intermediate steps in the construction of HAWC, such as HAWC with 30 tanks and 
with 100 tanks. We also investigated tanks that are instrumented with
as many as 7 PMTs per tank. In this PMT/tank range, the scaler
effective area scales as $N_{PMT}$, the total number of PMTs and/or tanks in the
detector.

\subsection{Background simulation for the scaler DAQ}
\label{sec:noisescaler}

We expect the total ($>1/4$ photoelectron) rate in the detector to be
$B=\unit[18]{MHz}$ ($\unit[20]{kHz} \times 900$). Even if the number of air
showers in given time intervals can be considered to follow a Poissonian
distribution, the distribution of the total noise rate $B$ is not Poissonian. This
is because some of the sources of noise are correlated. Sources that
produce two or more PMT signals in the detector result in a
distribution that is wider than a Poissonian because of \textit{double
  counting} in the measured average. In order, the  most important
sources of correlation are: (a) air showers resulting in more than one PMT
registering a signal, (b) PMT afterpulses and (c) Michel electrons due to muons
that stop inside HAWC tanks. The correlations result in a distribution that is
wider than a Poissonian. The Fano factor $F$ \cite{fano} describes how much wider the real
distribution is with respect to a Poissonian with the same mean value but no
correlation: $\sigma^2_B = F\;B$. Given a signal rate $S$, background 
rate $B$ and Fano factor $F$, then the significance of a given observation is: 
\begin{align}
\label{eq:SignificanceScaler}
\mathrm{Sigf} &= \frac{S\, \Delta T}{\sqrt{F \Delta T B}} \nonumber \\
              &= \sqrt{\frac{\Delta T} {F N_{PMT} 
                R_{PMT}}} \int^{E_{max}}_{E_{min}} \textup{d}E \;
                \frac{\textup{d}N}{\textup{d}E} A_{eff}^{scaler}(\theta),
\end{align} 
where $N_{PMT}$ is the number of PMTs in the detector, $R_{PMT}$ is the average
PMT rate, $\Delta T$ is the observation window and $\theta$ is the
zenith of the GRB being studied. Note that since the
background rate $B$ scales as $N_{PMT}$, and since the scaler
effective area also scales as $N_{PMT}$, then
Eq.~\ref{eq:SignificanceScaler} implies that the significance of
observations scales as $\sqrt{N_{PMT}}$. However the dependence of the
Fano factor on $N_{PMT}$ has not yet been studied.

During the operation of HAWC, the width of
the noise distribution, and hence the Fano factor, will be measured
experimentally. In the mean 
time, we have developed a dedicated background simulation to describe
the PMT noise rate and calculate the Fano factor. The objective of the
simulation is not to produce a series of distinct \textit{events} as 
customary in particle physics simulation, but to produce a PMT \textit{hit
stream} as measured by the scalers.

\begin{figure*}
  \centering
  \subfigure[Fano factor vs. Cutoff]
  {
    \label{fig:FanoVsLowECut}
    \includegraphics[width=0.45\linewidth]{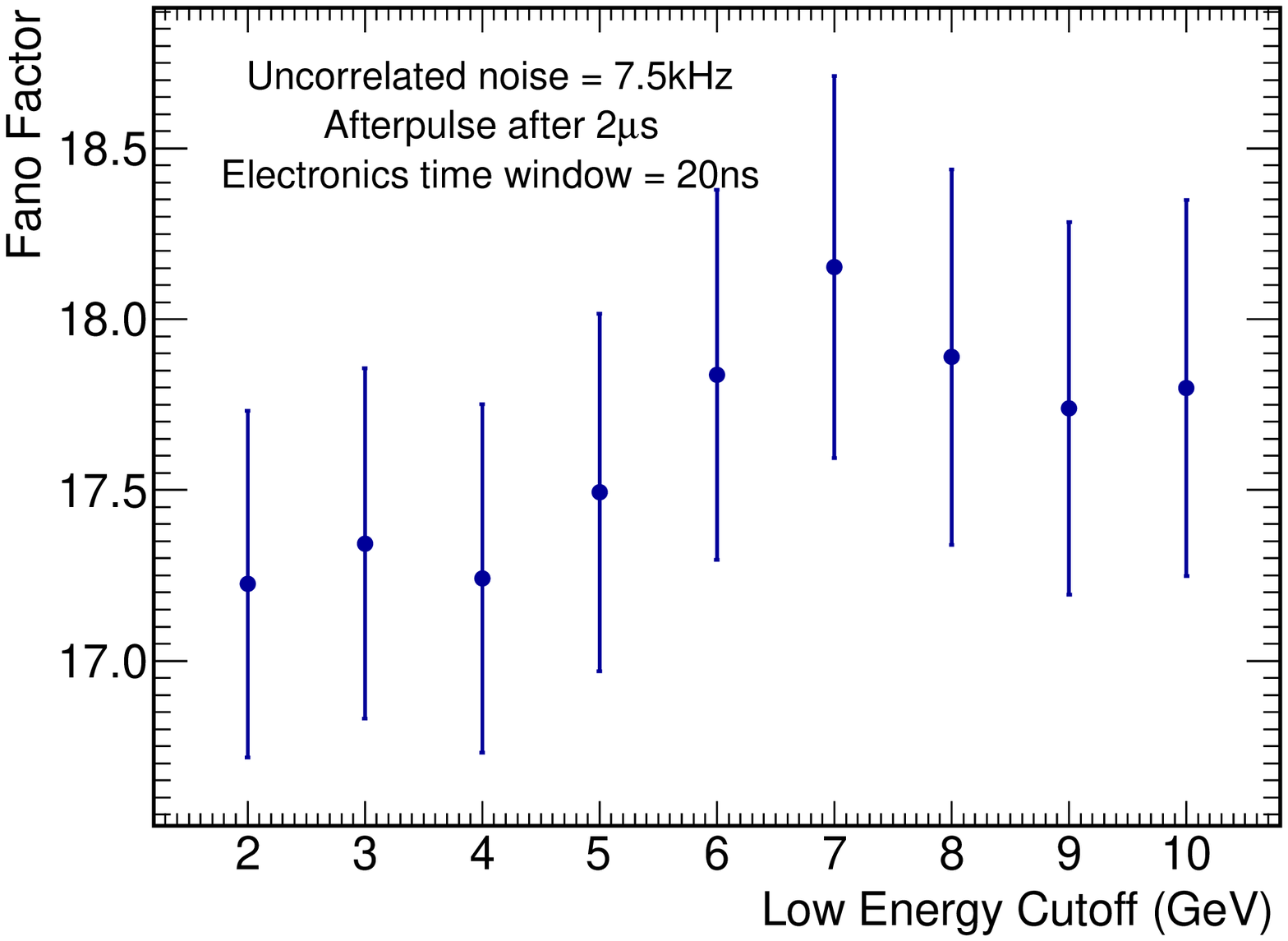}
  }
  \subfigure[Global PMT noise distribution]
  {
    \label{fig:FanoDistribution}
    \includegraphics[width=0.45\linewidth]{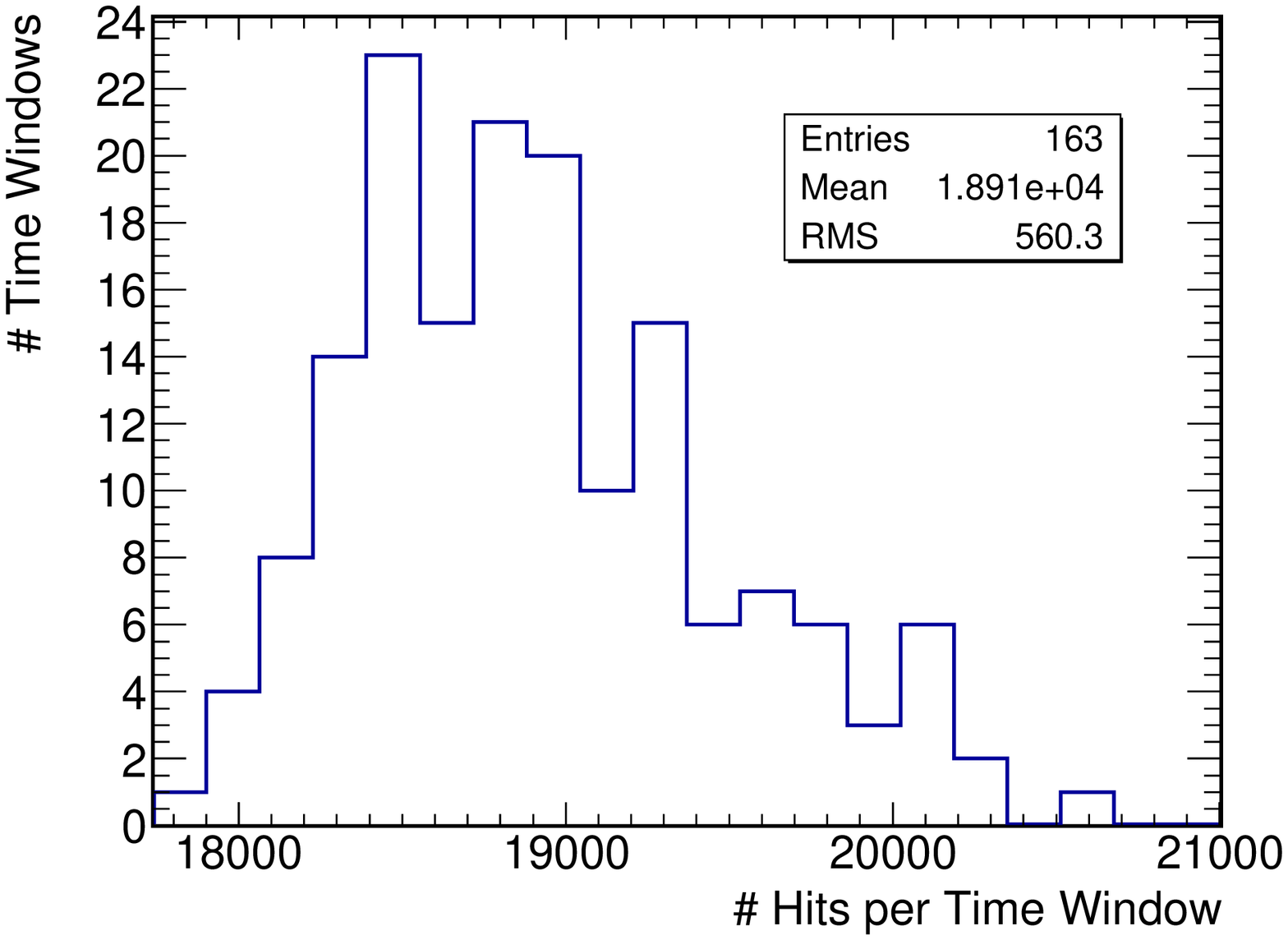}
  }
  \caption[Simulation of the noise rate of the HAWC scaler system]
          {{\bf Simulation of the noise rate of the HAWC scaler system}.}
  \label{fig:ScalerNoiseRate}
\end{figure*}

The simulation begins similarly to that of the main DAQ with a set of cosmic
ray events simulated with CORSIKA and a GEANT4-based detector simulation. We
assume an $E^{-2.7}$ power-law distribution for the primary cosmic rays and
normalize the rate with the ATIC measurements at high energies
($\approx$~\unit[100]{GeV}) \cite{ATIC}. Below \unit[10]{GeV} the cosmic ray spectrum is affected by the solar
modulation and the Earth's magnetic field. Because the value of the Fano factor is
almost independent of the shape of the spectrum at low energies
(Fig.~\ref{fig:FanoVsLowECut}) a detailed description of the spectrum is not
critical. We simulate the geomagnetic cutoff
($\approx$~\unit[8]{GeV} at the HAWC site) as a sharp cutoff, but again, this
choice is not critical. The result of this
simulation is a series of air showers each including a list of ideal
photoelectrons (i.e. not including the effects of electronics). The time of each air shower is
assigned 
randomly as expected from the assumed spectrum and the list of photoelectrons is
time sorted, i.e. a photoelectron stream is produced.

Uncorrelated Gaussian noise is added to the photoelectron
stream. This noise is expected to originate from thermal
noise, radioactive decays near or in the PMTs, etc. We considered
rates in the range of $\unit[5]{kHz}$ to $\unit[30]{kHz}$. We assume a default
value of $\unit[7.5]{kHz}$. If a higher rate is assumed the Fano
factor decreases, leading to almost the same sensitivity. Therefore,
for the ranges considered the choice of uncorrelated PMT noise is not critical.

The photoelectron stream is then further modified by adding PMT
afterpulses. The afterpulsing characteristics of HAWC PMTs are not known
yet, so we have used data for AMANDA (8 inch) and IceCube (10 inch)
PMTs as a reference. Each photoelectron is assigned a 6\% probability
of producing an afterpulse per photoelectron. 

The final step is a simplistic electronics simulation. The HAWC
front-end boards will enforce a minimum width for discriminated PMT
pulses 
of $\unit[20]{ns}$ (as a reference, the average 1 photoelectron pulse
at low threshold will be $\approx \unit[150]{ns}$). We 
merge all photoelectrons that are in coincidence in this time window and
call the resulting list a PMT hit stream. This electronics simulation
is conservative and will result in a larger Fano factor than a more
detailed simulation. Thus our crude electronics description does not
result in overestimating the scaler sensitivity. 

The final hit stream is then binned into the scaler time windows
(Fig.~\ref{fig:FanoDistribution}).
Using the default parameters we obtain a Fano factor of 17.4, which
reduces the sensitivity of HAWC scalers by a factor of 4.2 with respect
to assuming Poissonian PMT noise. 

\subsection{Sensitivity of the scaler system to GRBs}
\label{sec:sensitscalers}

We have used various spectra of the type $\textup{d}N/\textup{d}E \propto E^{\gamma}$
with sharp high-energy cutoffs to determine the sensitivity of the
scaler system to gamma-ray transients. A range of spectral indices $\gamma$
between -3 and -1 and a range of cutoffs between \unit[10]{GeV} and \unit[10]{TeV}
were tested, similar to those examined by the main DAQ. Again, effects
of the EBL are ignored.

\begin{figure*} 
  \centering
  \subfigure{
    \label{P:FermiFluence0}
    \includegraphics[width=0.45\linewidth]{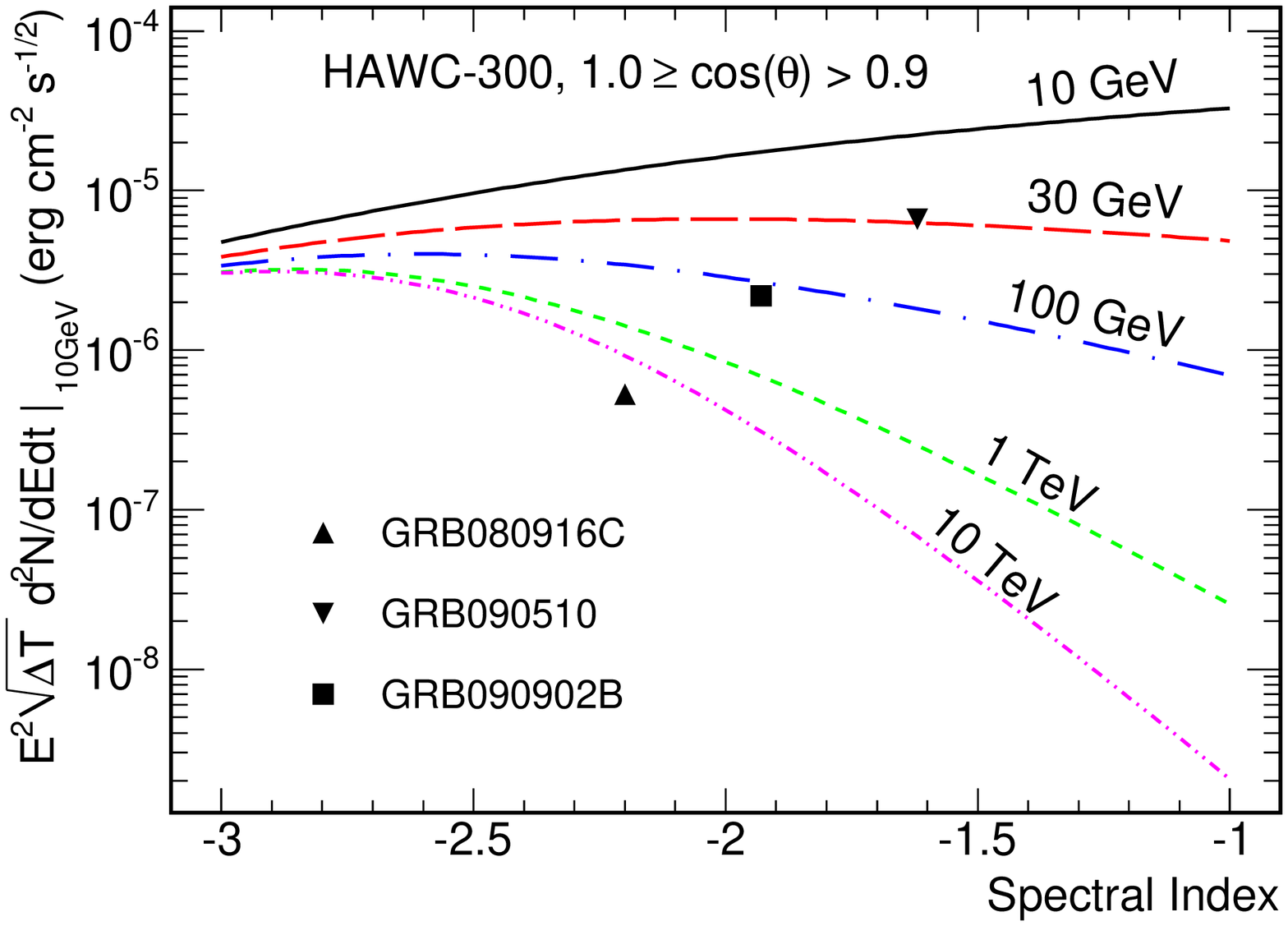}
  }
  \subfigure{
    \label{P:FermiFluence1}
    \includegraphics[width=0.45\linewidth]{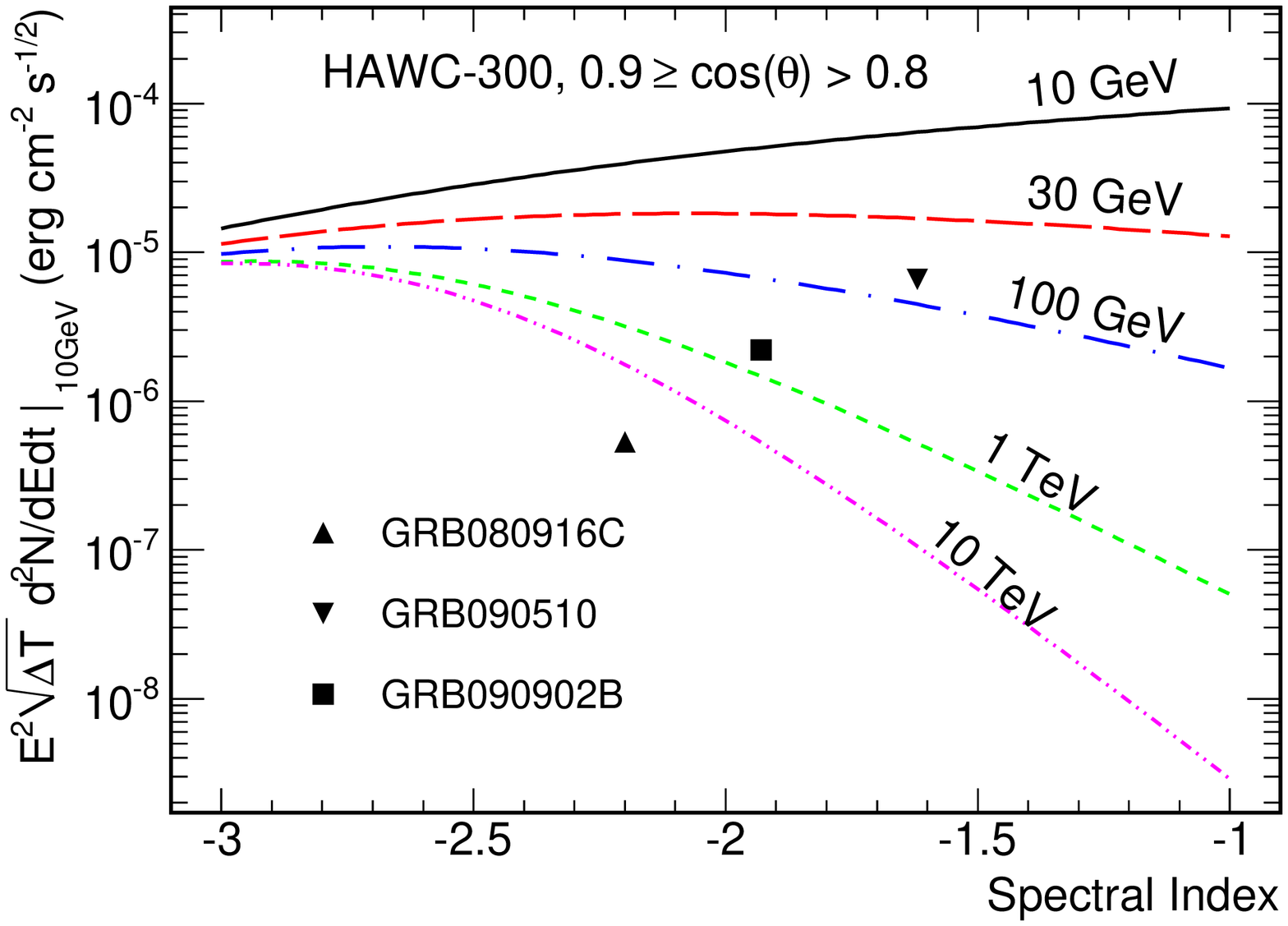}
  }
  \subfigure{
    \label{P:FermiFluence2}
    \includegraphics[width=0.45\linewidth]{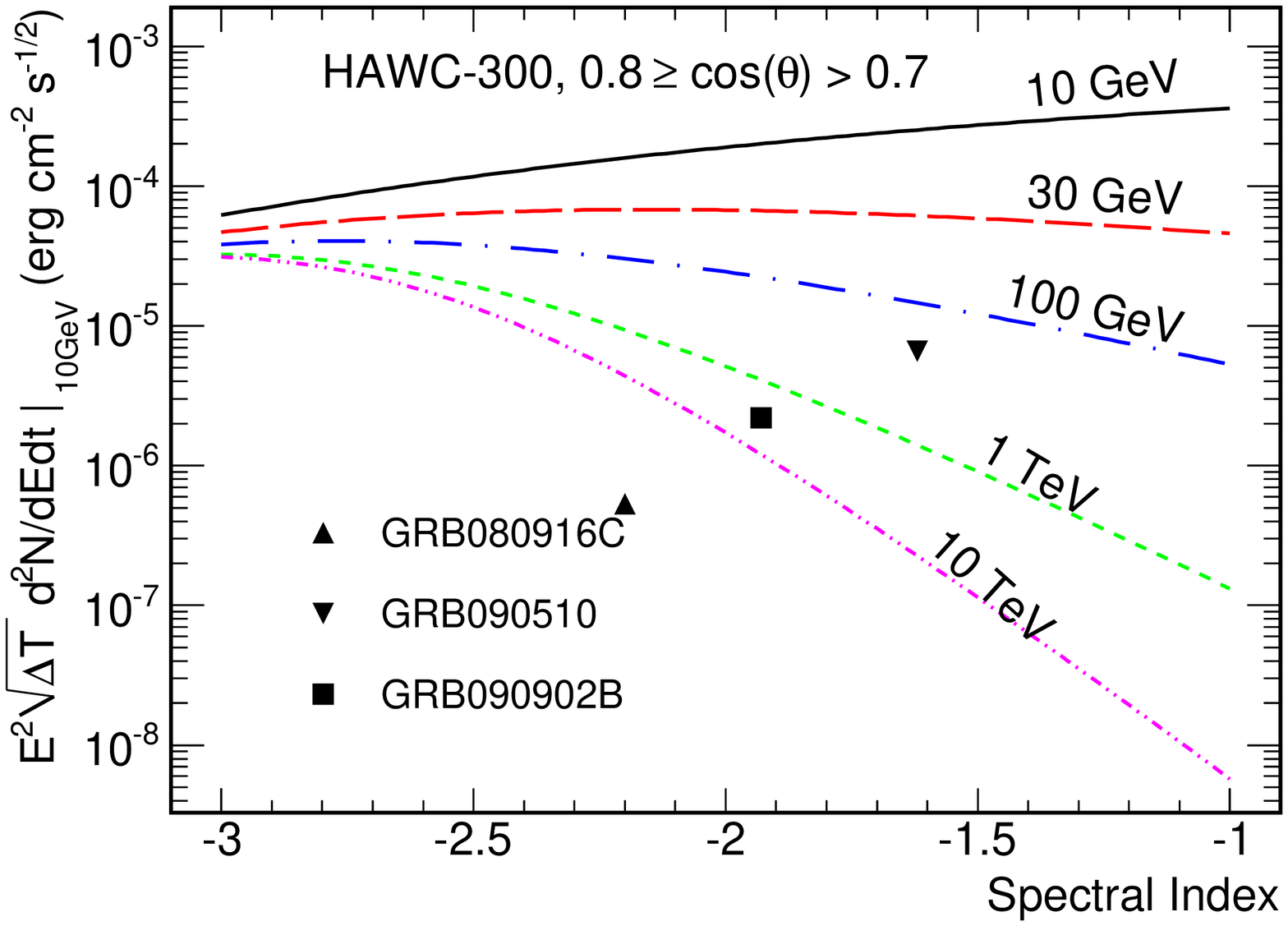}
  }
  \subfigure{
    \label{P:FermiFluence3}
    \includegraphics[width=0.45\linewidth]{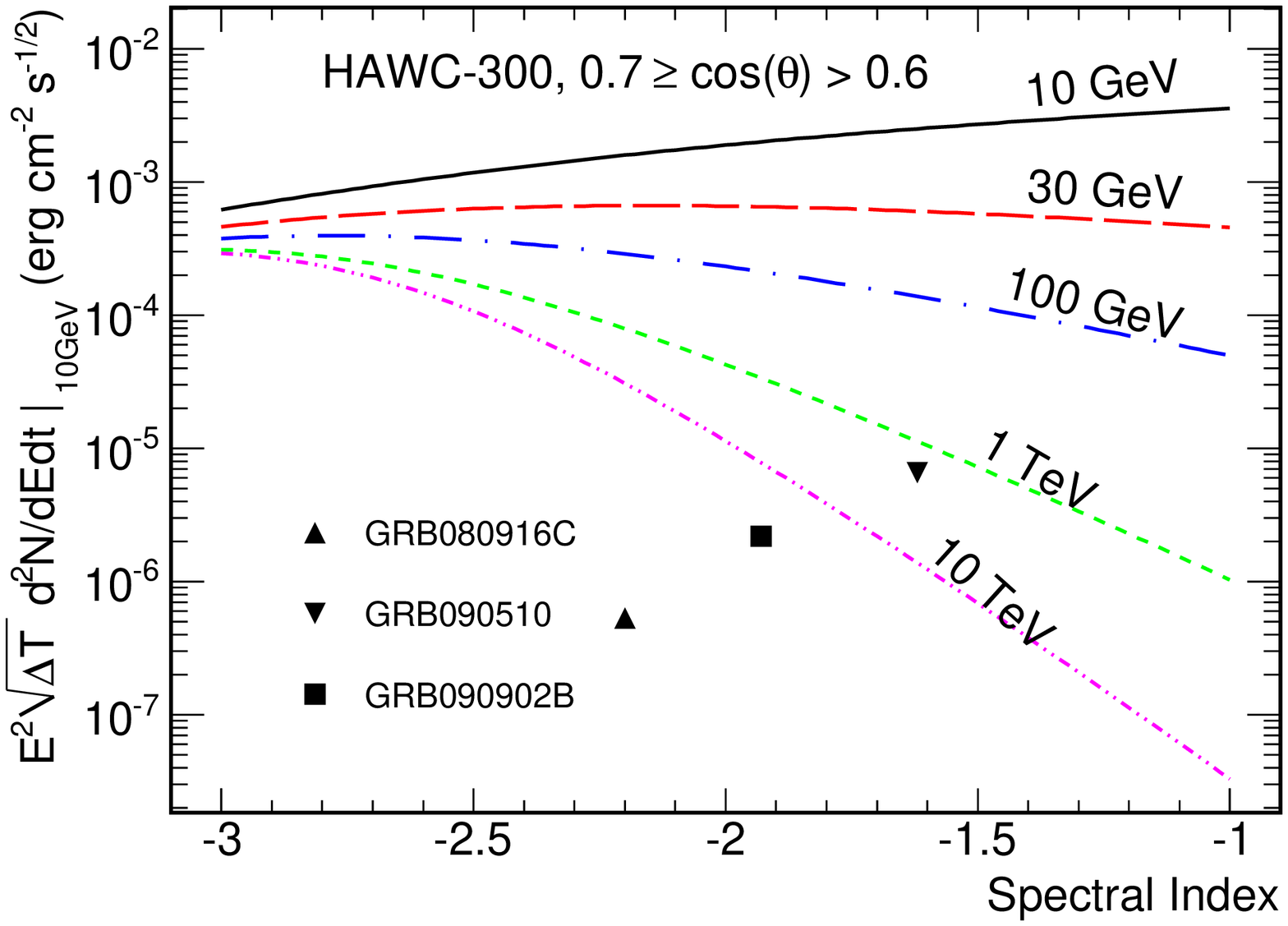}
  }
  \caption[Scaler sensitivity to GRBs]
    {{\bf Sensitivity of the scaler system to GRBs.}
       Necessary flux at \unit[10]{GeV} multiplied by the square
       root of the GRB duration to produce a 5~$\sigma$ signal in the
       HAWC scaler system. The scaler
       sensitivity is proportional to $\sqrt{\Delta T}$. We assume that the GRB spectrum includes a 
       power-law component with index $\gamma$ and a hard cutoff in addition 
       to the standard Band function.  At the energies to which HAWC is 
       sensitive, only the power-law component is relevant.The plots for five 
       different cutoff energies are shown in each picture. Each picture refers 
       to a specific zenith angle bin. Data from 3 different GRBs are inserted for 
       comparison \cite{grb080916c,grb090510,grb090902b}.
    }
  \label{fig:FermiFluence}
\end{figure*}

An interesting question is whether HAWC scalers are able to observe the same
GRBs that have been seen by Fermi LAT. Figure \ref{fig:FermiFluence} shows
HAWC's sensitivity, calculated with equation
\ref{eq:SignificanceScaler} compared to GRBs that have been detected
by Fermi LAT.
We conclude that the most promising cases for detection are GRBs that have a
non-Band hard power-law component such as GRB~090510 and
090902b. Fermi LAT observations of these two GRBs were made up to 
\unit[30]{GeV} without any indication of a cutoff. Therefore HAWC
scalers would observe GRBs similar to 090510 if they were to happen again within a zenith range
of $0-26^\circ$ (equivalently $0.9 < \cos\theta < 1.0$ or \unit[0.065]{sr}). If high-energy
emission from GRBs extends beyond \unit[30]{GeV}, then HAWC scaler observations
become even more significant while Fermi LAT struggles due to limited physical
size. If high-energy emission extends up to \unit[100]{GeV}, the HAWC scalers can make
observations in a field of view of \unit[0.125]{sr} ($0-37^\circ$ zenith range).

Also interesting is to compare the sensitivity of HAWC to other
detectors that use the single particle techinique. Because of size,
high altitude, better design (optically isolated water Cherenkov
detectors, as opposed to single ponds, scintillators or resistive
plate chambers), HAWC will be at least one order of magnitude more
sensitive than Milagro \cite{milagroScalers}, Pierre Auger
\cite{augerScalers},  LAGO \cite{lago}, ARGO
\cite{argoScalers}, etc. Our
simulations show that even VAMOS using scalers is slightly more
sensitive than Milagro. Note that even though LAGO is at a
significantly higher altitude and has a similar design to HAWC, it is
significantly smaller.

\section{Joint sensitivity of main DAQ and scaler DAQ to GRBs}

At a basic level the information provided by both the main DAQ and scaler system
will be a measure of the strength of the signal, which corresponds to
the number of events for the main DAQ and the significance of the
observation for the scaler DAQ. The main DAQ allows for event by event
energy reconstruction. Above $\approx \unit[1]{TeV}$ the energy can
be reconstructed from the data taken with the triggered DAQ
system. However at energies below $\approx \unit[1]{TeV}$, the showers
detected by HAWC are those in which the first interaction happens
lower than average in the atmosphere. At first glance, the difficulty
of HAWC to reconstruct energy in the range relevant to GRBs, may seem
as a drawback. However observations of GRBs by HAWC can be combined
with other detectors that have different energy response leading to
constrains in the spectrum. Furthermore, the energy responses of
scalers and the main DAQ are different. Their combined observations
(or lack of) can be used to constrain the very high energy spectrum.

\begin{figure*}
  \centering
  \subfigure{
    \includegraphics[width=0.45\linewidth]{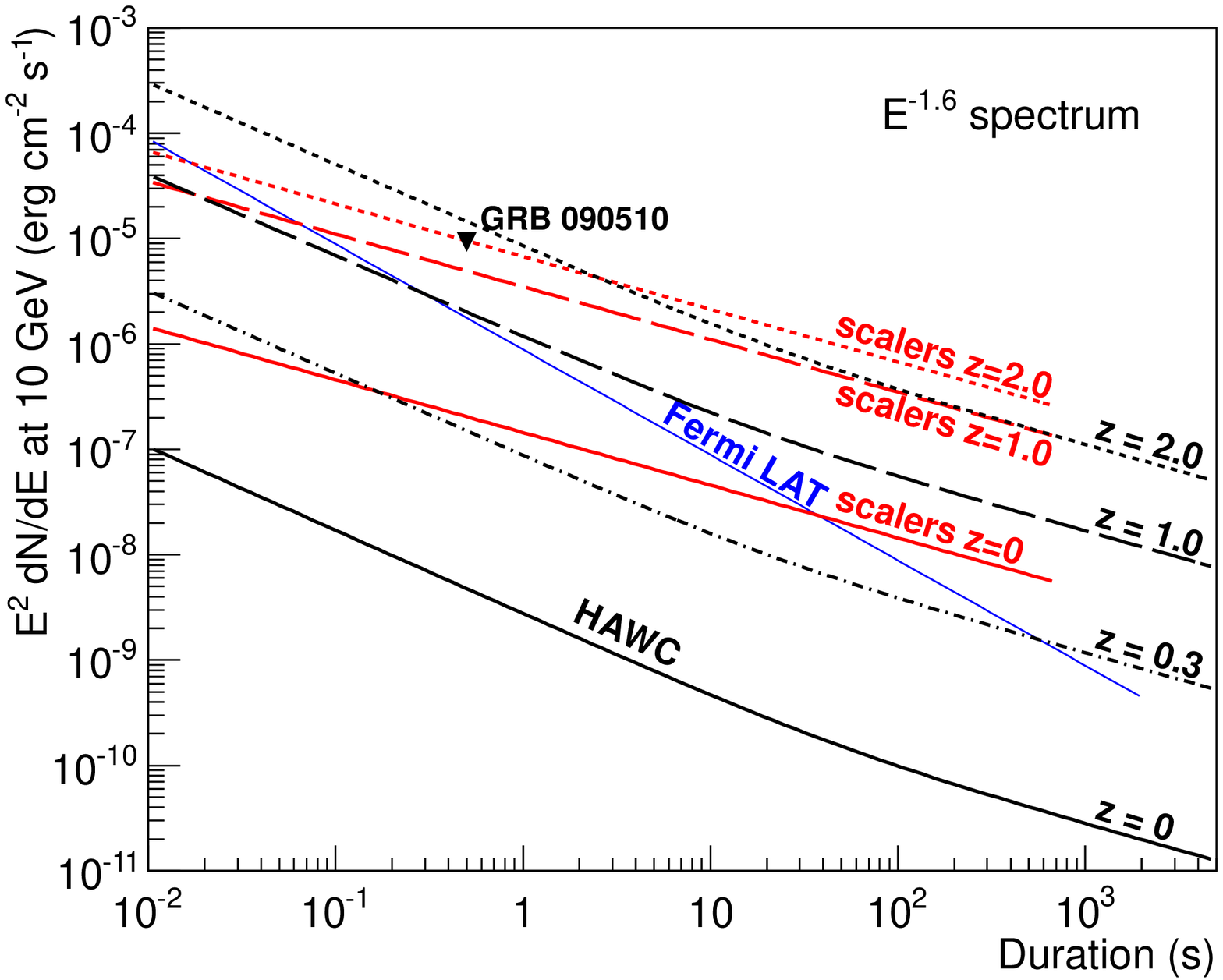}
    \label{fig:joint1}
  }
  \subfigure{
  \includegraphics[width=0.45\linewidth]{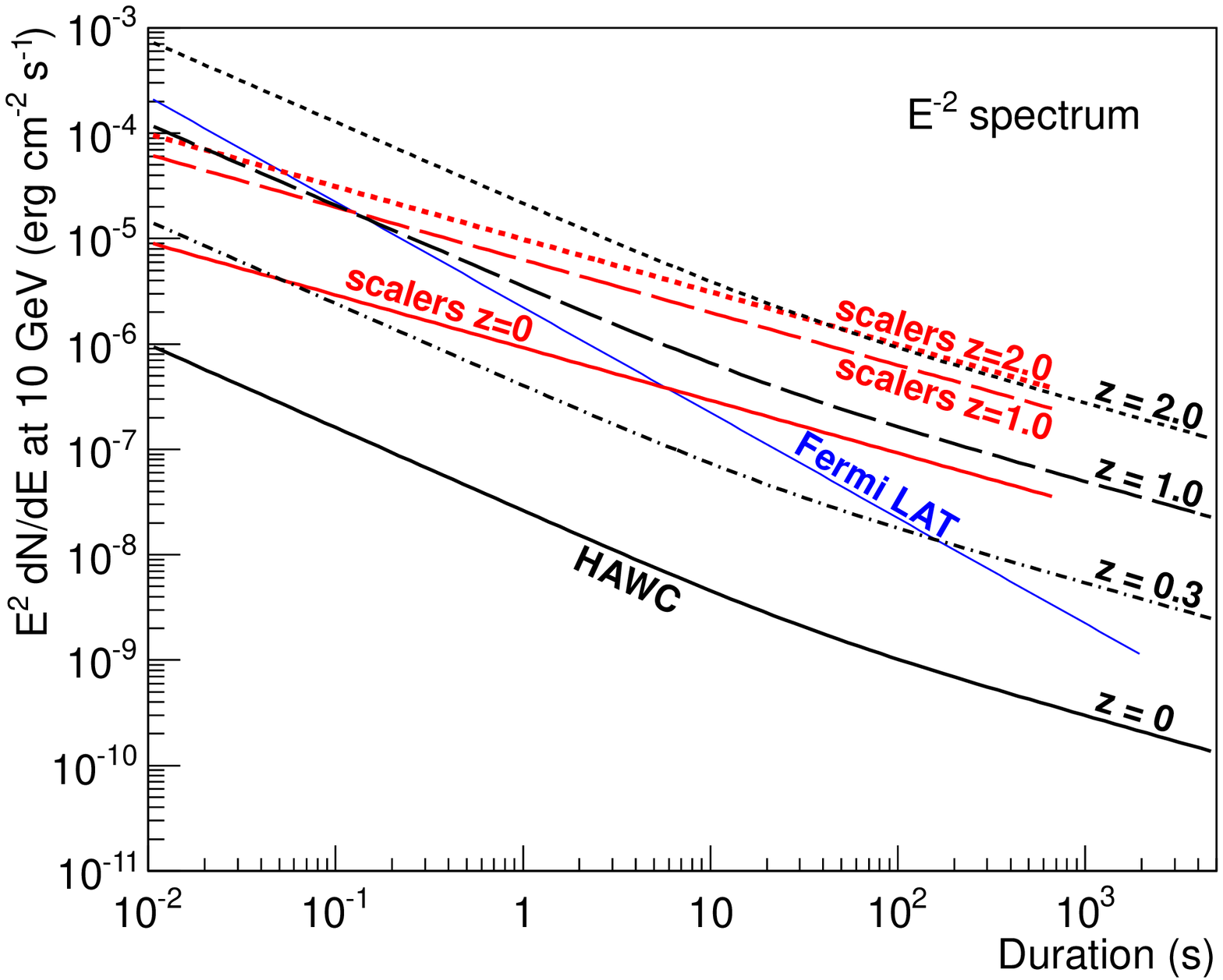}
    \label{fig:joint2}
  }
  \caption[Joint Sensitivity of main DAQ and Scalers for different GRB Durations]
  {{\bf Sensitivity of HAWC using the main DAQ and scalers as a
      function of burst duration.} The main DAQ uses a simple
    multiplicity trigger of 70 PMTs or more. The source position is set at a zenith angle of $20^\circ$.
  The source spectrum is $E^{-1.6}$ and $E^{-2.0}$ for the left and
  right plots respectively. 
  The Gilmore model of gamma ray attenuation by EBL \cite{gilmore09}
  is used to obtain the sensitivity curves for different redshifts.
  The lines for the scalers reflect the $5\, \sigma$ detection level.
  For the main DAQ the lines define the $5\, \sigma$ discovery potential.
  Also shown is the flux necessary for the observation of 1 photon
  above \unit[10]{GeV} by Fermi LAT.
  A marker is inserted in the left plot for GRB~090510 \cite{grb090510}.}
  \label{fig:joint}
\end{figure*}

The sensitivity of the scaler and main DAQs to GRBs are complementary:
the scaler DAQ covers a lower energy range and is able to provide information on
sudden increased rates whereas the main DAQ can reconstruct the energy and direction of
events at somewhat higher energy. Both systems will help provide information
on the spectra of GRBs. Figure \ref{fig:joint} shows the minimum flux required
to make a $5\, \sigma$ detection of a transient source with a zenith angle 
of $20^\circ$ and a duration between
$10^{-2}$ and $\unit[5\times10^{3}]{s}$ for both scalers and the main
DAQ. The trigger used by the main DAQ is a simple multiplicity trigger of 70 PMTs hit. Also
shown is the sensitivity of Fermi LAT assuming that at least one $\unit[>10]{GeV}$
photon is detected. The figure includes the effects of EBL as modeled by Gilmore
et al. \cite{gilmore09} at various values of redshift. Precisely because the
energy response of the scalers and main DAQ are different, the effects of EBL are
different for both techniques. The effects of a softer spectrum are less
pronounced on the scaler system. For all durations the sensitivity of scalers is
proportional to $1/\sqrt{\Delta T}$ as described in section
\ref{sec:sensitscalers}. For the main DAQ, the time dependence of flux
sensitivity is more complicated. At long durations, the background is large
enough so that the sensitivity scales as $1/\sqrt{\Delta T}$. At short durations
the background is very small and the sensitivity of the main DAQ is roughly proportional
to $1/\Delta T$.

We assume that the high-energy spectrum of a GRB may be described by
three parameters: the flux normalization at, in our case, $\unit[10]{GeV}$, the spectral index and
a high-energy cutoff. The high-energy cutoff is a quantity that is
particularly interesting to measure as it provides information about 
the bulk Lorentz boost factor of the GRB jet, probes the EBL or provides
information about the highest energy to which GRBs accelerate
particles. Depending on whether various satellites or ground based
instruments study a given GRB, then various parameters of the high-energy
spectrum can be constrained or measured.

As an example of how external information and a joint scaler-main DAQ
analysis could be performed we use a sample GRB. This GRB is
assumed to have similar, but not identical, characteristics to
GRB~090510. We use
\begin{equation}
  \frac{\textup{d}N}{\textup{d}E} = C\, \left( \frac{E}{\unit[1]{GeV}} \right)^\gamma
\end{equation}
and assume $\textup{d}N/\textup{d}E$ at $\unit[10]{GeV}$ to be 
$\unit[0.58]{GeV^{-1}\; m^{-2}\; s^{-1}}$ (or $E^2 \textup{d}N/\textup{d}E$
$= \unit[9.32 \times 10^{-6}]{erg\; cm^{-2}\; s^{-1}}$), a high energy spectral
index of $-1.6$ and a Heaviside high-energy cutoff at $\unit[150]{GeV}$. We
set the zenith angle to 20$^\circ$. EBL absorption is ignored because the 
objective of this example is to illustrate
how both systems in combination can measure a spectral cutoff. Using
the simulations described above, we obtain 17 expected events for the sample GRB 
seen by the main DAQ and $\approx$ 200,000 PMT hits over typical background
for the scaler system, leading to an $11\, \sigma$ significance by the
scaler system for this reference GRB.

\begin{figure}
  \centering
  \includegraphics[width=0.5\linewidth]{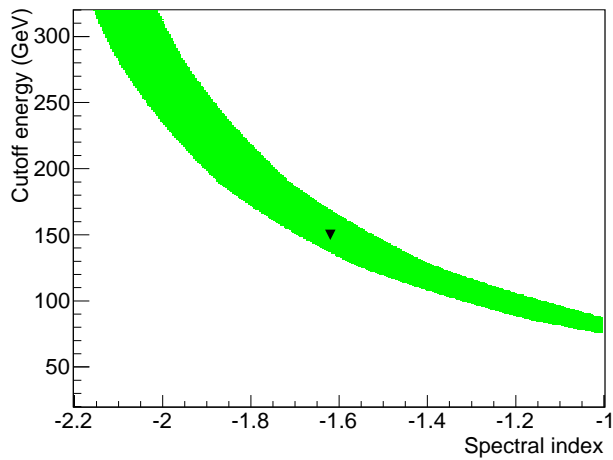}
  \caption[Joint Sensitivity of main DAQ and Scalers for different Energy Cutoffs]
  {{\bf{A simulation of a measurement constraining
  the spectral index and cutoff of a GRB.}} 
  The simulation combines the information from the main DAQ (with trigger threshold $nHit > 70$) 
  and  the scalers.
  The simulated GRB spectrum has a spectral index of $-1.62$
  and a Heaviside cutoff at $\unit[150]{GeV}$ (shown by black triangle).
  The flux normalization, chosen to match GRB~090510,
  was used to set a significance level in both DAQ systems.
  The green band shows the region where the flux values
  derived from the observations by the two systems
  agree within $\pm$25\%.
  Systematic errors are not included.}
  \label{fig:jointCut}
\end{figure}

We now assume that the main (scaler) DAQ system delivered the number of events
(significance) computed above and test the ability of
a combined analysis to constrain the spectral parameters.
In order to find the region of the parameter space consistent with the measurements,
the simulation described above is repeated for various 
values of a hypothetical Heaviside high-energy cutoff and 
spectral index.
For each point on the spectral index -- cutoff plane,
the computed number of events (significance) is used to find
the value of $\textup{d}N/\textup{d}E$ at $\unit[10]{GeV}$
which corresponds to the measured number of events (significance).
The values delivered by the main DAQ and scalers
are then compared to find the region where they are consistent.
The allowed region,
assuming a 25\% error bar between the two measurements,
is shown in Fig.~\ref{fig:jointCut}.
As can be seen, the combined analysis yields
meaningful constraints on the GRB spectrum.
The allowed region could be further narrowed down
by including data from other experiments (e.g. Fermi)
and theoretical constraints.
It should be noted that the values of
$\textup{d}N/\textup{d}E$ at $\unit[10]{GeV}$,
obtained with scalers, vary by $\approx \pm$40\%
throughout the allowed region.
Within this error bar, the measurement of the flux normalization
is independent of the spectral parameters measurement.
A similar technique was employed in \cite{atkins03}
to constrain spectral parameters of GRB~970417a.

In this example we have taken into account the statistical
uncertainties of the signal expectation in the main DAQ
system, shown as a band in Fig.~\ref{fig:jointCut}. However
the statistical uncertainty in the scaler system is 
extremely small. Therefore the uncertainty of the measurement by the
scalers is dominated by still unknown
systematic uncertainties. The variation of PMT noise rates due to upper
atmospheric conditions (which in turn affect air shower development),
local weather and instrumental effects are expected to be the main
sources of systematic uncertainties in the scaler system.

This example demonstrates that through the operation of the
scaler system and the main DAQ, HAWC will be able to extend the
sensitivity to high-energy emission by GRBs to energies currently
inaccessible to Fermi LAT, while performing prompt observations. In
particular it will be able to make measurements of a high-energy
spectral cutoff. Future studies with real GRBs will use more detailed
information and uncertainties provided by external sources such as sky
localization, redshift, spectrum, etc. 

\section{Scientific prospect and conclusions}

HAWC, a ground based EAS detector, will have the
capability of detecting GRBs at high energies. The
simulations presented in this paper show that HAWC will be able to
detect GRBs with characteristics similar to those of some of the brightest GRBs
seen by Fermi LAT. In particular we have shown that if bursts such as
GRB~090510 were to repeat with a zenith angle of
$26^\circ$ or less they would be detected with $5\, \sigma$ or greater significance. HAWC
will be particularly useful for GRBs with hard non-Band power-law
high-energy spectral components. Two methods for detection of GRBs
will be used by HAWC. The main DAQ will acquire and reconstruct data
on a shower by shower basis. The scaler system will search for a
statistical excess on the combined noise rate of all PMTs. As opposed
to Fermi LAT, with a fixed physical size, the effective area of both
scalers and main DAQ increases with energy.
Thus either of the explained
detection methods will expand upon the energy sensitivity of
current detectors. Also, because HAWC is a wide field of view detector
with near 100\% duty cycle, it will be able to make GRB observations
in the prompt phase. An example of how to perform a joint study of
GRB spectra using both scalers and the main DAQ was shown. HAWC,
in unison with satellite or ground based detectors, will be able to
measure the high-energy GRB components including a possible high-energy
cutoff. Important astrophysical information will be deduced from
spectral cutoffs such as the bulk Lorentz boost factor of GRB jets, the
effects of the EBL and the maximum energy to which GRBs accelerate
particles.

\section*{Acknowledgements}
This work has been supported by: the National Science Foundation, the
US Department of Energy Office of High-Energy Physics, the LDRD
program of Los Alamos National Laboratory, Consejo Nacional
de Ciencia y Tecnolog\'{\i}a (grants 55155, 103520, 105033, 105666,
122331 and 132197), Red de F\'{\i}sica de Altas Energ\'{\i}as, 
DGAPA-UNAM (grants IN105211, IN112910 and IN121309, IN115409),
VIEP-BUAP (grant 161-EXC-2011) and the University of Wisconsin Alumni 
Research Foundation.

\end{document}